\documentclass[a4paper]{article}
\usepackage[english]{babel}
\usepackage[utf8]{inputenc}

\usepackage{booktabs}
\usepackage{tabu}

\usepackage[a4paper,top=3cm,bottom=2cm,left=3cm,right=3cm,marginparwidth=1.75cm]{geometry}

\usepackage{amsmath}
\usepackage{amssymb}
\usepackage{graphicx}
\usepackage{csquotes}

\usepackage{latexsym}

\usepackage[style=chem-acs, articletitle=true, sorting=none]{biblatex}
\usepackage{multirow}

\usepackage[hang,small,bf]{caption}
\usepackage{tikz}
\usetikzlibrary{backgrounds,shadows.blur,fit,decorations.pathreplacing,shapes}
\newcommand{\ket}[1]{\ensuremath{\left|#1\right\rangle}} 
\newcommand{\bra}[1]{\ensuremath{\left\langle#1\right|}} 

\title{Generalized unitary coupled cluster excitations for multireference molecular states optimized by the Variational Quantum Eigensolver}
\author{Gabriel Greene-Diniz \\
        David Mu\~{n}oz Ramo\\
        Cambridge Quantum Computing Ltd.\\
        9a Bridge Street, CB2 1UB Cambridge\\
        United Kingdom}
\date{\today}

\begin{document}
\maketitle
\begin{abstract}
The variational quantum eigensolver (VQE) algorithm, designed to calculate the energy of molecular ground states on near-term quantum computers, requires specification of symmetries that describe the system, e.g. spin state and number of electrons. This opens the possibility of using VQE to obtain excited states as the lowest energy solutions of a given set of symmetries. In this paper, the performances of various unitary coupled cluster (UCC) ans\"{a}tze applied to VQE calculations on excited states are investigated, using quantum circuits designed to represent single reference and multireference wavefunctions to calculate energy curves with respect to variations in the molecular geometry. These ans\"{a}tze include standard UCCSD, as well as modified versions of UCCGSD and \textit{k}-UpCCGSD which are engineered to tackle excited states without undesired spin symmetry cross-over to lower states during VQE optimization. These studies are carried out on a range of systems including H$_2$, H$_3$, H$_4$,  NH, and OH$^{+}$, CH$_2$, NH$^{+}_{2}$, covering examples of spin singlet, doublet and triplet molecular ground states with single and multireference excited states. In most cases, our calculations are in excellent agreement with results from full configuration interaction calculations on classical machines, thus showing that the VQE algorithm is capable of calculating the lowest excited state at a certain symmetry, including multireference closed and open shell states, by setting appropriate restrictions on the excitations considered in the cluster operator, and appropriate constraints in the qubit register encoding the starting mean field state. 
\end{abstract}

\section{Introduction}

The theoretical elucidation of the energy and properties of molecules and materials at the atomistic level has been identified for a long time as one of the most straightforward applications of quantum computing. This assessment is motivated by the relatively low requirements of the corresponding quantum circuits in comparison with other applications \cite{nisq_preskill}. The use of quantum computers for atomistic simulations promises to bring radical advances in a variety of fields including pharma, photovoltaics, aeronautics, electronics and energy generation, among others. Most research efforts have been focused on efficient algorithms for the calculation of the molecular ground state energy, although in recent years work has been performed in the calculation of other quantities like excited state energies \cite{linear1, vqd, parrish_mcvqe, veiss_ch2, ollitrault19} or energy derivatives \cite{obrien_derivatives, mitarai_derivatives, parrish_derivatives}. 

Current quantum computers display high levels of noise at different stages of their operation, and it is expected that near-term ((i.e. noisy intermediate scale quantum (NISQ)) machines will continue having to grapple with this problem \cite{nisq_preskill}. Motivated by this fact, a variety of hybrid quantum-classical algorithms based on the Variational Quantum Eigensolver (VQE) \cite{peruzzo} have been proposed over the years. VQE's main advantage relies on its improved resiliency to noise and reduced circuit depth compared with pure quantum algorithms like quantum phase estimation. Since VQE's inception, many improvements have been proposed: schemes to simplify and generalize the wavefunction ansatz \cite{lee_kupcc, adapt_vqe, ms_troyer_ucc, ucc_google, ryabinkin_ucc}, techniques to reduce the impact of noise in the quantum circuit \cite{stabilizer_vqe} or in the classical minimization step \cite{ryabinkin_penalties, scikit_quant} or strategies to reduce the number of measurements of the hamiltonian \cite{h_measure1, alphaVQE_prl}, among others. A comprehensive review of these techniques may be found in Refs. \cite{aspuru_review, mcardle_review}.

VQE is designed to calculate the energy of molecular ground states, as the variational nature of the algorithm drives the calculation to the minimum of the energy function. The calculation typically starts by encoding a mean-field wavefunction obtained by a classical calculation with techniques like the Hartree-Fock method. In the second quantization formalism, the qubit register represents the occupations of the molecular spin orbitals. However, encoding the mean-field wavefunction into a qubit register is associated with defining some symmetries that describe the molecular system, such as the number of electrons or the spin state. One could perform experiments in which there is a selection of symmetries corresponding to states different from the ground state, for example setting the spin state of the qubit register to a triplet in the hydrogen molecule. These symmetries would constrain the wavefunction to drive the VQE algorithm towards local energy minima that satisfy them, effectively transforming this algorithm into a tool to determine the lowest excited state compatible with a given symmetry. 

However, studying excited states with VQE leads to a possible complication. Molecular excited states usually display higher entanglement than ground states and a mean-field trial wavefunction is a bad starting point for a correlated calculation on these systems. In these cases, a multireference initial state is needed to properly describe entanglement in these systems. It is well documented in the literature that conventional coupled cluster methods fail to obtain the appropriate description of multireference states \cite{szalay_cc_review, bartlett_cc}, and need to be adapted for these cases. In this sense, it is interesting to investigate the performance of VQE with the standard UCCSD ansatz in the calculation of these kinds of problems. An FCI approach to studying multireference states on quantum computers has also been proposed \cite{sugisaki19}, yet due to the reliance on quantum phase estimation is likely not to be suitable for the NISQ era. Recently, advanced ans\"{a}tze have been proposed which have been engineered to tackle the calculation of states with strong multireference character \cite{lee_kupcc, adapt_vqe}. These ans\"{a}tze allow for excitations in the cluster operator which do not distinguish between occupied and virtual orbital subspaces (so-called generalized excitations), and thus the various determinants in the multireference expansion are treated in an even-handed way by the cluster operator. The accuracy of these ans\"{a}tze in the case of excited states is interesting to assess, particularly in relation to the above-mentioned issues of symmetry preservation of the state during wavefunction optimization. 

In this paper, we explore these interesting topics by applying VQE to a range of molecules with different geometries and electronic structures. This range covers various degrees of excited state quasi-degeneracy, as well as different choices for the starting Hartree-Fock wavefunction and the correlation ansatz. The simplest system analyzed is H$_2$, which is representative of molecules with a closed shell singlet ground state and a triplet excited state. For the doublet case, we choose H$_3$ in a linear geometry as the simplest system to analyze. We then study the methylene radical, CH$_2$, whose ground state is a spin triplet with two unpaired electrons and its lowest excited states are spin singlets with strongly entangled character. CH$_2$ is well known as a good model for the study of electron correlation in excited states \cite{green_ch2_exp, slipchenko_ch2}. In the context of quantum computing, it has already been used as a test molecule for adiabatic state preparation and phase estimation techniques \cite{asp_ch2, veiss_ch2}. We follow the lead of these studies and propose CH$_2$ as a good test case for the study of entanglement in quantum chemistry for quantum computing.

Following CH$_2$, other diradical molecules are then studied, including NH$^{+}_{2}$, NH, and OH$^{+}$, to show the generality of our results in the context of multireference excited states. Finally, the bond-angle dependence of the ground and excited states of H$_{4}$ is studied using some of the ans\"{a}tze applied to the diradical molecules. H$_4$ a useful model to study the effect of a continuous change in the strength of quasi-degeneracy as one of the H-H bonds is broken to dissociation \cite{paldus_H4_93, li_paldus_os-cc95}, and it is used here as an attempt to explain some features of the previous calculations of more realistic molecules. As we will show in this paper, allowing for fully unrestricted single excitations in generalized UCC ans\"{a}tze can lead to deterioration of the symmetry of the desired excited state during VQE optimization. We show how to prevent the unwanted transition to symmetries of lower energy while preserving the benefits of generalized ans\"{a}tze, in particular the ability to access multireference states, thus boosting the effectiveness of generalized UCC ans\"{a}tze for calculating excited states.

While the calculation of lowest energy states for a given set of symmetry quantum numbers using variational methods is well established in classical approaches to quantum chemistry, the novelty of this work lies in the fact that much of the performance and behavior of variational approaches to excited states has yet to be tested in a quantum computational context. For example, ans\"atze that are appropriate or easy to implement for quantum methods are not necessarily appropriate for classical methods, or vice versa, and so the performance (i.e. whether desired symmetries are retained) of ans\"atze used in classical methods may not necessarily carry over to a quantum computational setting, especially quantum computational approaches to excited molecular states. A key example of this is the implementation of unitary coupled cluster (UCC) operator in a circuit based quantum algorithm \cite{romero18}. UCC cannot easily be implemented in classical approaches due to the lack of termination of the Baker-Campbell-Hausdorff expansion of the similarity transformed Hamiltonian. This problem is not present in the quantum implementation of UCC, as unitary operations are naturally mapped to quantum gates that can be applied to qubits in a quantum circuit. However, the UCC operator cannot be directly implemented as a quantum circuit, but has to be approximated via a Trotter decomposition, which introduces errors that may have an impact on the conservation of symmetries of the calculation. In addition, as classically implementable CC performs poorly for strongly correlated multireference excited states, it is interesting to investigate the ability of carefully adapted UCC ans\"atze to capture these states for different molecules. We note that previous work in this area has claimed that unwanted symmetry breaking is a general feature of the variational quantum eigensolver (VQE) applied to molecular states, and variational penalties to the objective function will in general be needed when applying VQE to states that are not the molecular ground state \cite{ryabinkin_penalties}. We show that with appropriate UCC ans\"atze, such penalty factors for retaining the spin symmetry or electron number are not always needed when accessing excited or ionized states. In fact, the undesired ionization of the H$_{2}$ molecule during variational optimization of the neutral excited state previously reported \cite{ryabinkin_penalties} may be a by-product of the use of so-called hardware efficient ans\"atze in which the preparation of the reference state is itself variationally parametrized \cite{ryabinkin_penalties, lio2_dimer}, which implies removing symmetry constraints from the calculation. Other works propose modified ans\"{a}tze based on symmetry-conserving quantum gates to explore these kinds of states \cite{economou_sym}. Thus, by demonstrating the ability of variational methods to capture a range of multireference excited states of various molecules in a quantum computational setting, our work contains novel contributions to the field of quantum computation applied to chemistry simulations.

The structure of this paper is described as follows. In section \ref{methods} we explain the methodology used for the calculations, which includes a brief overview of the $k$-UpCCGSD ansatz and the modifications to it used in this work. Following this, section \ref{results} presents the results obtained in this work, beginning with the H$_2$ and H$_3$ molecules. Next, the results for diradicals with triplet ground states are presented, focusing first on CH$_2$ which is used as a test case to demonstrate the pitfalls of generalizing all single excitations with no restrictions, and the modifications we use to avoid these pitfalls. This is followed by a report of the results for H$_{4}$, in which the square geometry is deformed into a chain via a trapezoid.  We finalize our study with a discussion of the results obtained, and draw conclusions relevant to the study of multireference molecular excited states using unitary coupled cluster run on quantum computers.

\section{Methods} \label{methods}

All the calculations have been performed using our EUMEN quantum chemistry package in combination with the ProjectQ quantum simulator \cite{projectq} and auxiliary functions provided by the OpenFermion library \cite{openfermion}. Our calculations required the computation of molecular integrals to define the second quantization hamiltonian. This step was done with the Psi4 package \cite{psi4}, which was also used to perform Full Configuration Interaction (FCI) and classical CCSD and EOM-CCSD calculations for comparison purposes. Classical optimization of UCC excitation amplitudes was performed using the L-BFGS-B optimization protocol of the SciPy Python library \cite{scipy}. 

We used different basis sets and active spaces depending on the species considered. For H$_2$ and H$_3$, we used the 6-31G basis set in order to be able to correlate the high-spin wavefunctions. This basis is prohibitively expensive to use in a simulator for the diradical molecules and for H$_{4}$. Therefore, we resorted to a STO-3G basis set for these cases. We also froze the 1s core orbital of CH$_{2}$, NH$^{+}_{2}$, NH, and OH$^{+}$ to reduce the size of the qubit register needed for the simulations. This setup is enough to capture the effects of correlation at different spin states for these molecules.

\begin{table}
\centering
\begin{tabular}{cc} 
\hline
Molecule & Initial wavefunction qubit register \\
\hline
\hline
\multirow{2}{7em}{H$_2$ (6-31G)} & $\ket{11000000}$ (S)\\
& $\ket{10100000}$ (T) \\
\hline
\multirow{2}{7em}{H$_3$ (6-31G)} & $\ket{111000000000}$ (D)\\
& $\ket{101010000000}$ (Q) \\
\hline
\multirow{4}{7em}{NH$^{+}_{2}$ and CH$_2$ (STO-3G)} & $\ket{111110100000}$ (T) (only for CH$_2$) \\ 
& $\ket{111111000000}$ (S1) \\ 
& $\frac{1}{\sqrt{2}}\ket{111111000000} - \frac{1}{\sqrt{2}}\ket{111100110000}$ (S2) \\ 
& $\frac{1}{\sqrt{2}}\ket{111101100000} - \frac{1}{\sqrt{2}}\ket{111110010000}$ (S3)\\
\hline
\multirow{4}{7em}{NH and OH$^{+}$ (STO-3G)} & $\ket{1111110000}$ (S1) \\
& $\frac{1}{\sqrt{2}}\ket{1111110000} - \frac{1}{\sqrt{2}}\ket{1111001100}$ (S2) \\ 
& $\frac{1}{\sqrt{2}}\ket{1111011000} - \frac{1}{\sqrt{2}}\ket{1111100100}$ (S3)\\
\hline
\multirow{2}{7em}{H$_4$ (STO-3G)} & $\ket{11101000}$ (T)\\
& $\ket{11110000}$ (S1) \\ 
& $\frac{1}{\sqrt{2}}\ket{11110000} - \frac{1}{\sqrt{2}}\ket{11001100}$ (S2) \\ 
\hline
\end{tabular}
\caption{Initial state qubit encoding for H$_2$, H$_3$, H$_4$, NH, OH$^{+}$, NH$^{+}_{2}$ and CH$_2$ at different spin symmetries. The alternating spin-up/spin-down convention is used for the spin orbital occupations encoded in the qubit registers. Letters between brackets refer to spin multiplicities: S (singlet), D (doublet), T (triplet), Q (quadruplet). For CH$_2$, three different spin singlets are considered: S1, S2 and S3.}
\label{table_regs}
\end{table}

In these calculations we used the VQE algorithm with various UCC ans\"{a}tze, adapted for closed shell and open shell configurations. For the single determinant calculations of the ground and excited states of H$_2$ and H$_3$ we used the regular UCCSD ansatz. Following this, the importance of generalized excitations is investigated for all other molecules. Such excitations refer to the lack of distinction of occupied and virtual orbital subspaces, such that virtual-virtual and occupied-occupied transitions are also included for single ($T_1$) and double ($T_{2}$) excitation operators. The use of such excitations goes back to the work of Nooijen \cite{nooijen20} and Nakatsuji \cite{nakatsuji20} in their studies of the use of single and double excitation operators for obtaining the exact wavefunction. In addition to the previously proposed ans\"atze \textit{k}-UpCCGSD and UCCGSD \cite{lee_kupcc} which include generalized excitations, we also study the effect of modifications of these ans\"atze (discussed below) and apply these operators to a range of geometry points for all other molecules. 

For the \textit{k}-UpCCGSD cluster operator, single excitations are fully generalized in the sense of no distinction between occupied and virtual subspaces, while double excitations are restricted to electron pairs in the same spatial orbital (and these pair-double excitations are generalized). In an actual calculation, this cluster operator is applied $k$ times to the reference wavefunction, where each $k$ factor has variationally independent amplitudes,

\begin{equation} \label{eqn_k-upccgsd}
\ket{\Psi} = \prod_{k}(e^{T_{k} - T^{\dagger}_{k}})\ket{\Phi},
\end{equation}

\begin{equation} \label{eqn_k-up_T}
T_{k} = T_{1, k} + T_{2, k} .
\end{equation}

\noindent As mentioned above, calculations were also carried out with the UCCGSD ansatz \cite{lee_kupcc}, where the cluster operator retains the structure of the conventional UCC case, but all generalized single and double excitations are considered.

As we are interested in the interplay between spin symmetry and the different excitations present in the cluster operator, we consider single electron excitations in two different modes for all ans\"{a}tze. In the first mode, all general single excitations are individually considered, such that $\alpha$-$\alpha$ and $\beta$-$\beta$ spin transitions between a given pair of spatial orbitals $P$ and $Q$ have independent amplitudes. This can be expressed by writing

\begin{equation} \label{eqn_k-up_gs_T1}
T_1 = \sum_{P \neq Q}(t^{Q(\alpha)}_{P(\alpha)}\hat{a}^{\dagger}_{Q(\alpha)}\hat{a}_{P(\alpha)} + t^{Q(\beta)}_{P(\beta)}\hat{a}^{\dagger}_{Q(\beta)}\hat{a}_{P(\beta)})
\end{equation}

\noindent In the second mode, the independence of the $\alpha-\alpha$ and $\beta-\beta$ transitions is removed for single transitions, thus providing an extra constraint to the ansatz wavefunction during the variational procedure

\begin{equation} \label{eqn_k-up_gsprime_T}
T^{\prime}_1 = 2\sum_{P \neq Q}t^{Q}_{P}\hat{a}^{\dagger}_{Q}\hat{a}_{P} .
\end{equation}

\noindent Ans\"{a}tze constructed with this constraint have been labelled as UCCS$^{\prime}$D, $k$-UpCCGS$^{\prime}$D and UCCGS$^{\prime}$D.

In addition, we test constraining the cluster operator further by completely removing the single excitations altogether and retaining only the generalized pair doubles, in a manner reminiscent of seniority zero pair coupled cluster (pCCD) approaches previously explored in classical computational studies \cite{stein_sen0pCCD_14, bulik_SRCC_15}. The work of Scuseria \textit{et. al.} has shown that seniority zero pCCD holds promise for capturing sufficient static correlation to yield good approximations for doubly occupied FCI \cite{stein_sen0pCCD_14}. With regards to VQE, this approximation has the added benefit of a reduced number of parameters to optimize. We applied this approach to the excited states of CH$_{2}$, using modified versions of the UCCSD, $k$-UpCCGSD and UCCGSD ans\"{a}tze, which we labelled as UCCD, $k$-UpCCGD and UCCGD respectively. As an example, the UCCGD operator becomes


\begin{equation} \label{eqn_uccgd}
\ket{\Psi_{GD}} = e^{T_{GD} - T^{\dagger}_{GD}}\ket{\Phi},
\end{equation}

\begin{equation} \label{eqn_uccgd_T}
T_{GD} = \sum_{pqrs}t^{rs}_{pq}\hat{a}^{\dagger}_{s}\hat{a}^{\dagger}_{r}\hat{a}_{q}\hat{a}_{p} ,
\end{equation}

\noindent where lower case indices $p, q, r, s$ label occupied and virtual spin orbitals. In both cases of UCCGSD and UCCGD only a few points of the energy curve were calculated due to the large computational expense associated with fully generalized double excitations.

Doubles-only cluster operators are also related to so-called Brueckner doubles (BD) cluster approximation \cite{handy89, mizukami20}, in which a double-only operator is applied to a reference determinant of Brueckner orbitals, the latter being obtained by absorbing the $e^{T_1}$ operator into Hartree-Fock orbitals \cite{handy89}. A recent paper has demonstrated the satisfaction of the Brueckner condition by UCCD applied to a set of optimized orbitals, where the latter are obtained by an optimized unitary rotation applied to a Hartree-Fock state. In order to investigate the use of Brueckner orbitals for the doubles-only operators mentioned here, we approximate the scheme of Mizukami /textit{et al} by taking advantage of the availability of single excitations from FCI. We extract the FCI singles excitation coefficients and use them to generate a reference determinant of Brueckner-like orbitals 

\begin{equation}
\ket{\Psi_B} = e^{T_{1,FCI}}\ket{\Psi_HF} .
\end{equation}

\noindent To this determinant $\ket{\Psi_B}$ we then apply the UCCGD or $k$-UpCCGD operators to approximate unitary forms of the Brueckner doubles coupled cluster approximation. We apply these operators to the singlet $\tilde{a}^1A_1$ state of CH$_2$ as a further investigation of the role played by single excitations in the VQE optimization of this state. The findings are discussed in section \ref{discussion}.

In several cases, we explore further constraints to the spin symmetry in the VQE algorithm. In particular, we follow the approach of constrained VQE optimization and restrict the Hilbert space dimension such that only states of a particular symmetry are considered \cite{vqd}. By including a constraining term in the energy function $E_{VQE}$ that is optimized,

\begin{equation} \label{eqn_penalty}
E_{VQE} = E + p|\bra{\Psi}S^2\ket{\Psi}|^2
\end{equation}

\noindent where $p$ is a penalty factor (a value of 10 was used in these calculations), energies corresponding to a particular spin symmetry can be enforced. 

Encoding of the Hamiltonian and the wavefunction was performed using the Jordan-Wigner scheme. The Bravyi-Kitaev \cite{bk_paper} scheme was also tested for the ground and first excited state of CH$_{2}$ for $k$-UpCCGSD, and no significant difference was observed relative to the Jordan-Wigner scheme. Hence the Jordan-Wigner scheme was used throughout the results presented in the next section. 

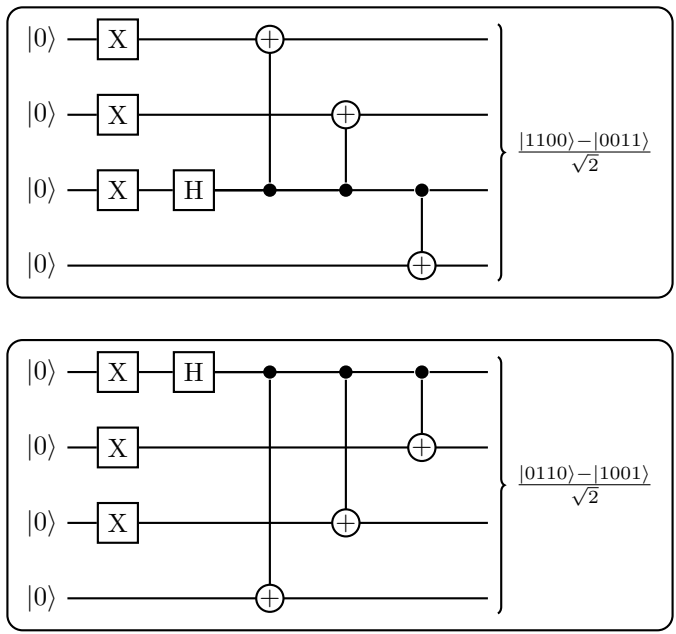
\begin{figure}[ht]
\begin{center}

\begin{tikzpicture}[thick]
\tikzstyle{circle}=[fill=none,shape=circle,minimum size=0.1cm,inner sep=0pt,outer sep=0pt,draw=black]
    \tikzstyle{operator} = [draw,fill=white,minimum size=1.5em] 
    \tikzstyle{phase} = [fill,shape=circle,minimum size=5pt,inner sep=0pt]
    \tikzstyle{surround} = [fill=white!10,thick,draw=black,rounded corners=2mm]

    \node at (0,0) (q0) {\ket{0}};
    \node at (0,-1) (q1) {\ket{0}};
    \node at (0,-2) (q2) {\ket{0}};
    \node at (0,-3) (q3) {\ket{0}};

    \node[operator] (op10) at (1,0) {X} edge [-] (q0);
    \node[operator] (op11) at (1,-1) {X} edge [-] (q1);
    \node[operator] (op12) at (1,-2) {X} edge [-] (q2);

    \node[operator] (op22) at (2,-2) {H} edge [-] (op12);

    \node[circle] (circle30) at (3,0) {+} edge [-] (op10);
    \node[phase] (phase32) at (3,-2) {} edge [-] (op22);
    \draw[-] (circle30) -- (phase32);

    \node[circle] (circle41) at (4,-1) {+} edge [-] (op11);
    \node[phase] (phase42) at (4,-2) {} edge [-] (op22);
    \draw[-] (circle41) -- (phase42);
    
    \node[circle] (circle53) at (5,-3) {+} edge [-] (q3);
    \node[phase] (phase52) at (5,-2) {} edge [-] (op22);
    \draw[-] (circle53) -- (phase52);
    
    \node (end1) at (6,0) {} edge [-] (circle30);
    \node (end2) at (6,-1) {} edge [-] (circle41);
    \node (end3) at (6,-2) {} edge [-] (phase52);
    \node (end4) at (6,-3) {} edge [-] (circle53);

    \draw[decorate,decoration={brace},thick] (6,0.2) to
	node[midway,right] (bracket) {\hspace{0.1cm}$\frac{\ket{1100}-\ket{0011}}{\sqrt{2}}$}
	(6,-3.2);
	
    \begin{pgfonlayer}{background} 
    \node[surround] (background) [fit = (q0) (q3) (bracket)] {};
    
    \end{pgfonlayer}
\end{tikzpicture}

\vspace{5mm}

\begin{tikzpicture}[thick]
\tikzstyle{circle}=[fill=none,shape=circle,minimum size=0.1cm,inner sep=0pt,outer sep=0pt,draw=black]
    \tikzstyle{operator} = [draw,fill=white,minimum size=1.5em] 
    \tikzstyle{phase} = [fill,shape=circle,minimum size=5pt,inner sep=0pt]
    \tikzstyle{surround} = [fill=white!10,thick,draw=black,rounded corners=2mm]
    %
    \node at (0,0) (q0) {\ket{0}};
    \node at (0,-1) (q1) {\ket{0}};
    \node at (0,-2) (q2) {\ket{0}};
    \node at (0,-3) (q3) {\ket{0}};

    \node[operator] (op10) at (1,0) {X} edge [-] (q0);
    \node[operator] (op11) at (1,-1) {X} edge [-] (q1);
    \node[operator] (op12) at (1,-2) {X} edge [-] (q2);

    \node[operator] (op20) at (2,0) {H} edge [-] (op10);

    \node[phase] (phase30) at (3,0) {} edge [-] (op20);
    \node[circle] (circle33) at (3,-3) {+} edge [-] (q3);
    \draw[-] (phase30) -- (circle33);

    \node[phase] (phase40) at (4,0) {} edge [-] (op20);
    \node[circle] (circle42) at (4,-2) {+} edge [-] (op12);
    \draw[-] (phase40) -- (circle42);
    
    \node[phase] (phase50) at (5,0) {} edge [-] (op20);
    \node[circle] (circle51) at (5,-1) {+} edge [-] (op11);
    \draw[-] (phase50) -- (circle51);
    
    \node (end1) at (6,0) {} edge [-] (phase50);
    \node (end2) at (6,-1) {} edge [-] (circle51);
    \node (end3) at (6,-2) {} edge [-] (circle42);
    \node (end4) at (6,-3) {} edge [-] (circle33);

    \draw[decorate,decoration={brace},thick] (6,0.2) to
	node[midway,right] (bracket) {\hspace{0.1cm}$\frac{\ket{0110}-\ket{1001}}{\sqrt{2}}$}
	(6,-3.2);

    \begin{pgfonlayer}{background} 
    \node[surround] (background) [fit = (q0) (q3) (bracket)] {};
    
    \end{pgfonlayer}
\end{tikzpicture}

\end{center}
\caption{Circuit diagrams for the preparation of the multiconfigurational states consisting of a linear combination of closed shell (top panel, register S2) or open shell (bottom panel, register S3) singlets for the diradicals and for H$_4$. The closed shell singlet corresponds to qubit register S2 from in Table \ref{table_regs}, while the open shell singlet corresponds to qubit register S3 in Table \ref{table_regs}. Using CH$_{2}$ as an example, the circuits act on the fifth, sixth, seventh and eighth qubits of the register.}
\label{lc-csS_circuit}
\end{figure}

As initial states, we first considered qubit registers encoding single reference open shell multiplets and closed shell singlets from a Hartree-Fock calculation created by applying X gates to the qubits representing the occupied Hartree-Fock molecular orbitals. Different charge states are also considered. For the CH$_2$ case, we considered two initial entangled states representing multireference open shell and closed shell singlet wavefunctions in addition to the closed shell singlet wavefunction constructed from a single Slater determinant. We will refer to them as S3, S2 and S1, respectively. We designed small quantum circuits to prepare these states, which we show in Figure \ref{lc-csS_circuit}. For CH$_{2}$ and NH$^{+}_{2}$, these circuits act on qubits 4 to 7 of the register, corresponding to the $sp$ hybridized valence shell and closest virtual orbitals. Qubits 0 to 3 are prepared as previously explained by applying X gates to represent the occupied spin orbitals below the valence shell. For the other diradical molecules, we consider only the low lying multireference excited states, thus only the `S' circuit registers are used for NH$^{+}_{2}$, NH, OH$^{+}$, specifically circuits S1, S2, and S3, which act on qubits representing the corresponding occupied valence orbitals. For H$_4$, we study the transition between the triplet and closed shell singlet using single reference and multireference wavefunctions when generalized UCC operators are used, during the dissociation of 2 H atoms as the square is deformed into a chain, which provides a good test of these UCC ans\"{a}tze under the continuous change of static correlation. Hence S1 and S2 are used for H$_4$. The details of the qubit registers encoding the wavefunctions for all molecules considered are shown in Table \ref{table_regs}.

\vspace{5mm}


\section{Results} \label{results}

We summarize here our experiments with VQE in which a range of molecular systems at several spin and charge states, including multireference states, are calculated using various unitary coupled cluster wavefunction ans\"{a}tze.  

\subsection{H$_2$ and H$_3$}

\begin{figure}[ht]
\begin{center}
\includegraphics[width=10cm]{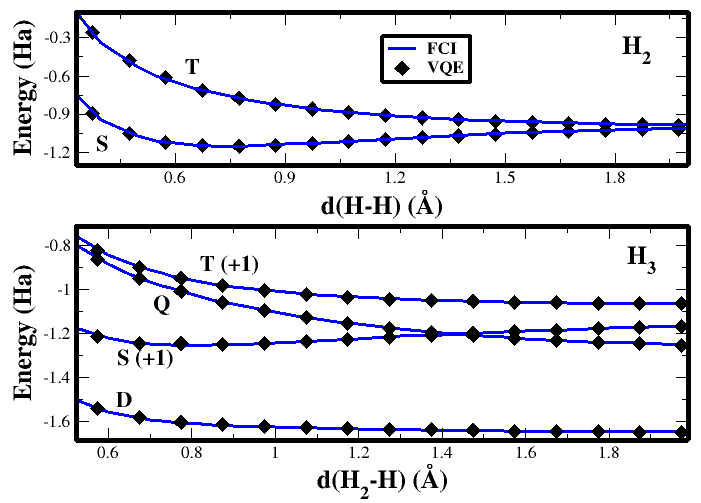}
\caption{Energy curves for the elongation of a H-H bond in H$_2$ (top graph) and linear H$_3$ (bottom graph) at different spin states and charge states using VQE, compared with classical FCI results. Singlet (S) and triplet (T) curves have been considered for H$_2$. In the case of H$_3$, states considered are doublet (D), quadruplet (Q), cation singlet (S+1) and cation triplet (T+1).}
\label{h2_curves}
\end{center}
\end{figure}

We begin our analysis with an evaluation of the energy curves of the H$_2$ molecule at different bond lengths. The ground state of this molecule is a spin singlet, and its first excited state has triplet symmetry. Our singlet calculation with the UCCSD ansatz, as expected, has a good match with the singlet energy curve calculated with classical methods. In addition, our calculation of the triplet state at the 6-31G level, using the T initial Hartree-Fock encoding shown in Table \ref{table_regs}, yields an energy curve that also matches faithfully the one obtained with FCI, as shown in Figure \ref{h2_curves} alongside the energy curve for the singlet ground state.

We study now the H$_3$ molecule with linear geometry and a 6-31G basis set. Here, we track the variation in energy with respect to the stretching of one H-H bond along the molecular axis. The spin states considered are the doublet (ground state with one unpaired electron) and the quadruplet (excited state with three unpaired electrons). Initialized qubit registers are shown in Table \ref{table_regs} for these spin multiplets. As in the H$_2$ case, the agreement between the VQE results using the UCCSD ansatz and the FCI and classical CCSD results is almost perfect at all distances considered (see Figure \ref{h2_curves}). An extension of this study to the cation H$_3^+$ in its singlet and triplet curves displays similar accuracy. No crossover is observed between energy curves in any of the states studied.

\subsection{CH$_2$} \label{ch2_subsec}

\begin{figure}[ht]
\begin{center}
\includegraphics[width=9cm]{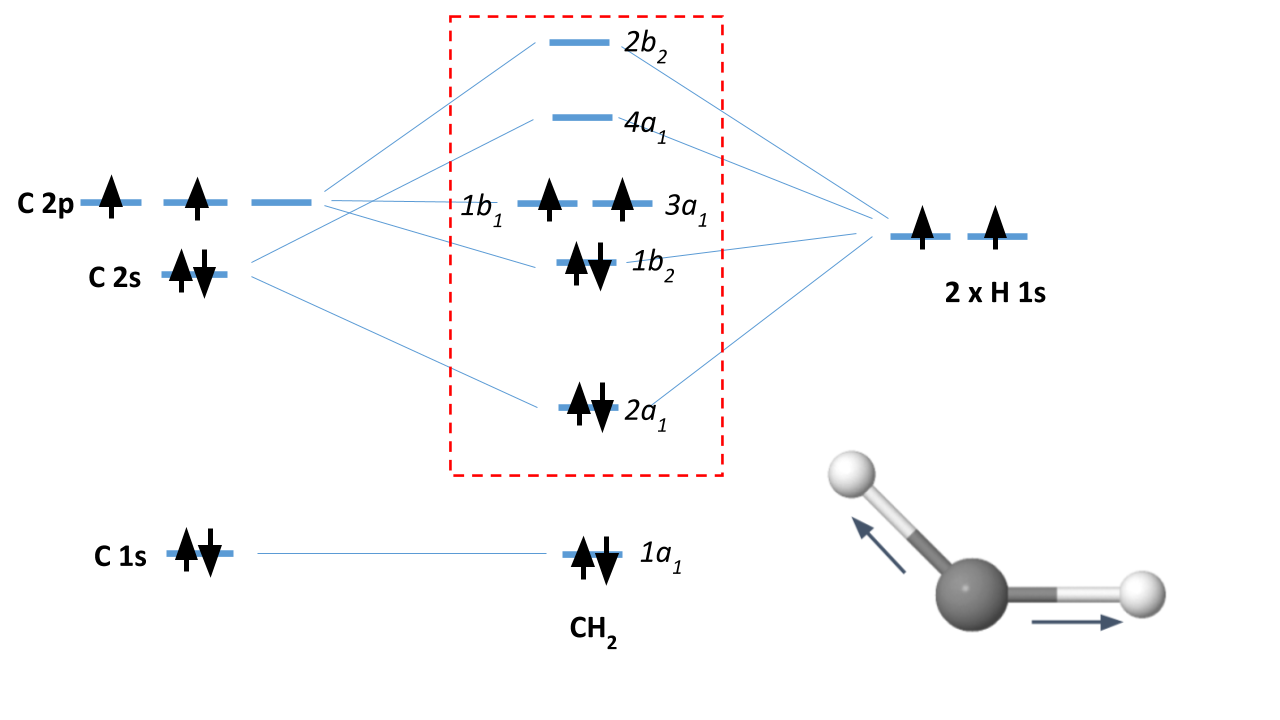}
\caption{Geometry and molecular orbital diagram for CH$_2$ in the ground state (triplet symmetry). The active space used in this work is indicated as the set of orbitals inside the red box.}
\label{mo_diagrams}
\end{center}
\end{figure}

The diradical CH$_2$ is used as the exemplary system to present the calculation of states with strong static correlation using generalized UCC excitations. In later sections, comparable results are also reported for other diradical molecules, followed by a final study in which trapezoidal H$_{4}$ is used as a toy system to explain some of the findings observed for the more realistic molecules. For now, we study the energy curves of CH$_2$ calculated with VQE for the symmetric stretching of both C-H bonds, fixing the H-C-H angle to 135$^{\circ}$, as shown schematically in Figure \ref{mo_diagrams}.
The ground state of CH$_2$ is a spin triplet with symmetry $\tilde{X}^3B_1$ whose main configuration is:

\begin{equation} \label{eqn_trip}
\tilde{X}^3B_1 \simeq (1a_1)^2 (2a_1)^2 (1b_2)^2 (3a_1)^1 (1b_1)^1  
\end{equation}

\noindent This configuration is shown in Figure \ref{mo_diagrams}. One can see that the highest occupied molecular orbital is doubly degenerate. The first excited state of this molecule is a closed shell singlet with multireference character and symmetry $\tilde{a}^1A_1$. A good description of this state is provided by the leading term: 

\begin{equation} \label{eqn_mr-sing}
\tilde{a}^1A_1 \simeq \lambda (1a_1)^2 (2a_1)^2 (1b_2)^2 (3a_1)^2  - \sqrt{1 - \lambda} (1a_1)^2 (2a_1)^2 (1b_2)^2 (1b_1)^2
\end{equation}
 
\noindent where $\lambda$ is a parameter that depends on the H-C-H angle of the molecule. The second excited singlet state is an open shell singlet with symmetry $\tilde{b}^1B_1$:
 
 \begin{equation}
\tilde{b}^1B_1 \simeq (1a_1)^2 (2a_1)^2 (1b_2)^2 (3a_1)^1\alpha (1b_1)^1\beta -  (1a_1)^2 (2a_1)^2 (1b_2)^2 (3a_1)^1\beta (1b_1)^1\alpha, 
\end{equation}
 
\noindent while the third lowest excited state is the closed shell singlet $\tilde{c}^1A_1$, which can be understood as a double excitation of the $\tilde{a}^1A_1$ singlet state:

\begin{equation}
\tilde{c}^1A_1 \simeq \lambda (1a_1)^2 (2a_1)^2 (1b_2)^2 (1b_1)^2  + \sqrt{1 - \lambda} (1a_1)^2 (2a_1)^2 (1b_2)^2 (3a_1)^2
\end{equation}
 
\noindent This ordering of the excited states has been reported in previous works on CH$_{2}$ \cite{green_ch2_exp, slipchenko_ch2}, and is also observed from classically computed FCI and equation of motion coupled cluster (EOMCC) energies. We have calculated energy curves for all these states using FCI and EOMCC classical methods in order to use them as useful benchmarks for our VQE results. We now analyze the results obtained for each initial qubit register with different methods. Results for the ground and excited states of CH$_{2}$ are shown in Figure \ref{ch2_curves_BL-only}.

\subsubsection{Triplet ground state calculation}
 
\begin{figure}[ht]
\begin{center}
\hspace{1.5cm}\includegraphics[width=8cm]{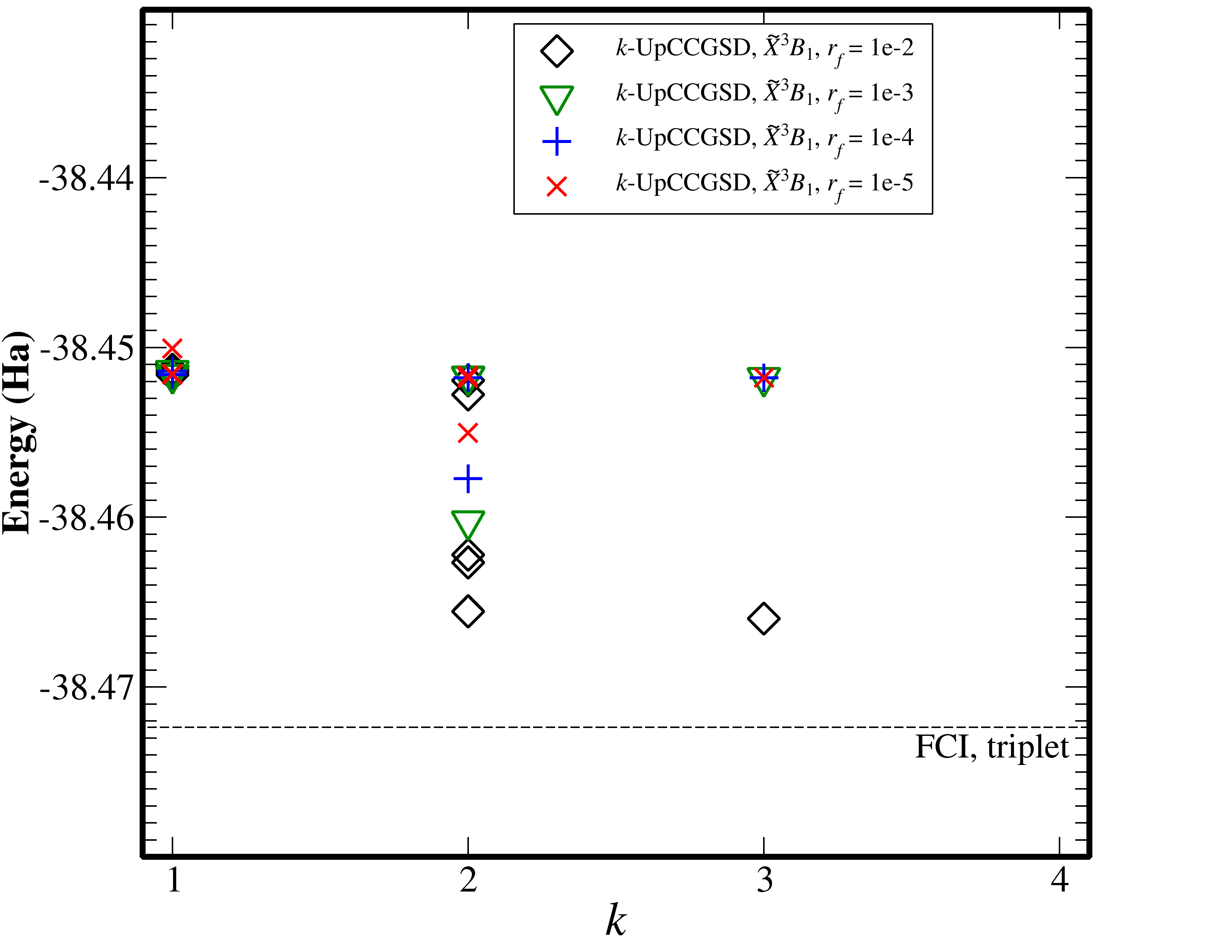}
\caption{Energy versus \textit{k} in \textit{k}-UpCCGSD applied to the triplet ground state of CH$_{2}$. The dashed line denotes the FCI energy. For \textit{k} $>$ 1, a spread in energies reached at the end of optimization is observed which depends on the variance of the initially randomized cluster amplitudes, given by the randomization factor $r_{f}$.}
\label{ch2_Evsk_T}
\end{center}
\end{figure}

\vspace{5mm}
 With the molecular integrals and qubit register for the initial ansatz set to the ground state triplet  $\tilde{X}^3B_1$ (qubit register T in table \ref{table_regs}), the FCI ground state energy was recovered by the VQE algorithm using the UCCSD and UCCGSD cluster operators.  
For the \textit{k}-UpCCGSD calculations, the ground state $\tilde{X}^3B_1$ energy at the equilibrium geometry was calculated as a function of \textit{k} up to \textit{k} = 3. As shown in Figure \ref{ch2_Evsk_T}, when the \textit{k}-UpCCGSD ansatz is adopted, variations of approximately 0.02 Ha in the optimized energy of $\tilde{X}^3B_1$ were observed when \textit{k} $>$ 1. It should be noted that the optimization procedure, which variationally optimizes the cluster amplitudes, begins with a randomization in which the initial amplitudes are Gaussian-distributed about 0 with a variance $r_{f}$. By plotting the optimized energies as a function of \textit{k} for repeated runs with a range of variances $r_{f}$ = $1\times10^{-5}$ - $1\times10^{-2}$, a spread in the optimized energies for \textit{k} $>$ 1 is observed. The spread depends on $r_{f}$, with lower energies (closer to the FCI energy) obtained for larger values of $r_{f}$ which permit sufficient variational flexibility for the \textit{k}-UpCCGSD ansatz to approach the FCI limit. 

\begin{figure}[ht]
\centering
\includegraphics[width=14cm]{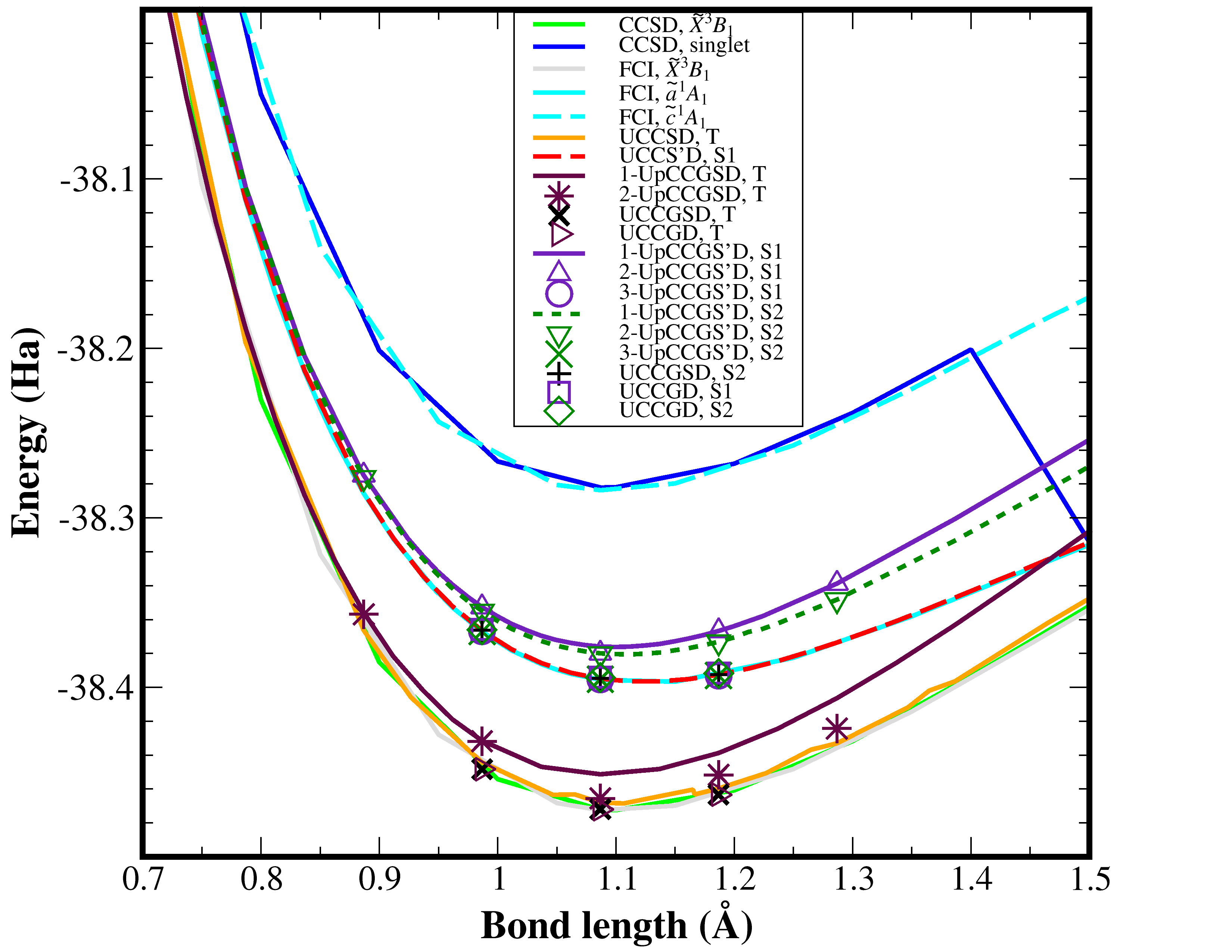}
\caption{Energy versus C-H bond length (PES) curves for the ground and excited states of CH$_{2}$, calculated using various representations of the cluster operator, and using different initial configurations of spin orbitals as input to VQE optimization. For \textit{k}-UpCCGSD (triplet) and \textit{k}-UpCCGS$^{\prime}$D (singlet), lines denote results for \textit{k} = 1, while the symbols of the corresponding color refer to results for \textit{k} $>$ 1 or fully generalized ans\"{a}tze. The black $\times$ and $+$ symbols refer to the initial state configurations T and S2 (used for the $\tilde{X}^3B_1$ and $\tilde{a}^1A_1$ states) respectively, calculated using UCCGSD. The square and diamond symbols refer to the UCC operator with fully generalized doubles and no singles. 
}
\label{ch2_curves_BL-only}
\end{figure}

This spread in energies for $\tilde{X}^3B_1$ using \textit{k}-UpCCGSD is associated with the presence of many unphysical local minima, which in turn can arise from the fact that \textit{k}-UpCCGSD energies are not invariant to unitary rotations of the spin orbitals \cite{lee_kupcc}. Hence, we follow the approach of Lee \textit{et al} \cite{lee_kupcc} and use the lowest energy solutions for the ground state. A decrease in the ground state energy (at the minimum of the PES) of 0.015 Ha is observed between \textit{k} = 1 and \textit{k} = 2. For the latter, the lowest energy solution matches that for \textit{k} = 3. This variation in ground state energy with respect to \textit{k} is in qualitative agreement with the calculations reported by Lee et al \cite{lee_kupcc}, who observed convergence with respect to \textit{k} for \textit{k} = 2 to within a few mHa for ground states calculated at the potential energy surface (PES) minimum using the minimal basis set STO-3G. Thus, for the \textit{k}-UpCCGSD calculations of the PES for ground and excited states of CH$_{2}$ we proceed with a maximum value of 2 for \textit{k}, while also presenting the results for \textit{k} = 1 for comparison, and adopt the randomization factor $r_{f}$ = $1\times10^{-2}$. Figure \ref{ch2_curves_BL-only} shows that the $\tilde{X}^3B_1$ energy is 0.02 Ha above the FCI limit near the minimum of the PES for \textit{k} = 1, hence \textit{k}-UpCCGSD is within 6 mHa of the FCI limit for the ground state when \textit{k} = 2. 

\subsubsection{Singlet excited states calculation} \label{ch2_singlets}

We focus now on the qubit registers encoding singlet functions, as the triplet ground state is well reproduced by CCSD, UCCSD, UCCGSD, and reasonably well for 2-UpCCGSD. The spin symmetries obtained following VQE optimization starting from registers S1, S2 and S3, using the different ans\"{a}tze considered in this work, are reported in Table \ref{spin_cross}. 

We start our discussion by studying the UCCSD case. Applying this ansatz without any kind of symmetry constraint to the S1 initial qubit register drives the VQE calculation to a spin crossover to the triplet ground state solution at all bond lengths considered. This is consistent with the variational principle underlying VQE. In contrast, using the UCCS$^{\prime}$D ansatz (with single excitation constraints) reproduces the FCI energy for the first excited state ($\tilde{a}^1A_1$) throughout the PES curve. 
Inspection of the wavefunction in the crossover case shows that the optimization process sometimes induces a change of sign in several leading amplitudes corresponding to single excitations from doubly occupied orbitals. These excitations drive the wavefunction to the S$_z$ = 0 component of the triplet state and produce the spin switching. This process is cancelled when single excitation constraints are applied. Interestingly, we find that conventional CCSD calculations with single excitation constraints on a classical machine fail to obtain the $\tilde{a}^1A_1$ state near the equilibrium geometry and converge instead to the $\tilde{c}^1A_1$ state, as shown in Figure \ref{ch2_curves_BL-only}. 

We use the \textit{k}-UpCCGSD and UCCGSD cluster operators next, and apply them to the S1 single reference register as well as the S2 multireference register. When unconstrained $k$-UpCCGSD is applied to S1 or S2 (representing closed shell singlets), we observe a lowering of the energy towards that of the ground state. To understand this behaviour, we performed an analysis of the expectation value of the total spin operator with respect to the optimized wavefunction $\bra{\Psi}S^2\ket{\Psi}$ for these states. In particular, for \textit{k}-UpCCGSD applied to the S1 qubit register, values of $\bra{\Psi}S^2\ket{\Psi}$ = 1 and $\bra{\Psi}S^2\ket{\Psi}$ = 2 were observed for \textit{k} = 1 and \textit{k} = 2 respectively, after optimization, indicating a mixture of singlet and triplet states and a crossover to the ground state triplet, respectively. When \textit{k}-UpCCGSD is applied to S2, no spin crossover ($\bra{\Psi}S^2\ket{\Psi}$ = 0) is observed for 1-UpCCGSD, while for 2-UpCCGSD the total spin expectation value is consistent with a triplet $\bra{\Psi}S^2\ket{\Psi}$ = 2 and the ground state $\tilde{X}^3B_1$ energy is reproduced. 

Interestingly, for unconstrained UCCGSD the effect of spin crossover towards the triplet only occurs when the initial state is prepared with the S1 register, whereas the S2 register retains its spin symmetry after VQE optimization using UCCGSD.  Thus, using a multireference initial wavefunction appears to be sufficient to prevent spin crossover and recover the FCI limit of the first excited state using UCCGSD. On the other hand, our results indicate the susceptibility of the \textit{k}-UpCCGSD ansatz to spin crossover for the closed shell single reference and multireference states. 

As the transition between singlet and triplet states involve single electron excitations \cite{slipchenko_ch2, krylov_o-shell_17}, we next study the effects of modifications in the single excitation amplitudes on spin crossover behaviour. First, we remove the generalization of single excitations to occupied and virtual subspaces, while retaining the independence of $\alpha-\alpha$ and $\beta-\beta$ transition amplitudes for a given pair of spatial orbitals. This results in the simplified ansatz ``$k$-UpCCSGD''. This ansatz is no longer applicable to multireference states, but provides a test case for investigating the origin of spin crossover. As shown in Table \ref{spin_cross}, no difference is seen in the spin crossover behaviour compared to $k$-UpCCGSD, ruling out the generalized nature of the single excitations as the origin of spin crossover.

Next we consider the modified ansatz $k$-UpCCGS$^{\prime}$D (see equations \ref{eqn_k-upccgsd} and \ref{eqn_k-up_gsprime_T}), which is applicable to closed shell singlet configurations. As can be seen from Table \ref{spin_cross}, this ansatz is robust to closed shell singlet-triplet spin crossover, thus providing a promising approach to calculate the first excited state of the diradical molecules. Another alternative to this ansatz (not shown for brevity), in which the single excitations were no longer generalized but single excitations are spin restricted as in equation \ref{eqn_k-up_gsprime_T}, i.e. ``$k$-UpCCS$^{\prime}$GD'', was also investigated for the S1 configuration. In this case the spin crossover is also prevented, as in $k$-UpCCGS$^{\prime}$D, but the accuracy relative to FCI is reduced for given values of $k$ (for $k$ up to $k$ = 5, the energy equates to the $k$ = 1 value of $k$-UpCCGS$^{\prime}$D). Thus, removing the generalization of the singles operator has the effect of worsening the convergence of the energy with respect to $k$. We find that values of $k$ too large to be tractable are required for achieving the FCI limit for $k$-UpCCS$^{\prime}$GD, and combined with its non-applicability to multireference excited states, we chose to omit this version of the ansatz from the energy curves. 

\begin{figure}[ht]
\centering
\includegraphics[width=14cm]{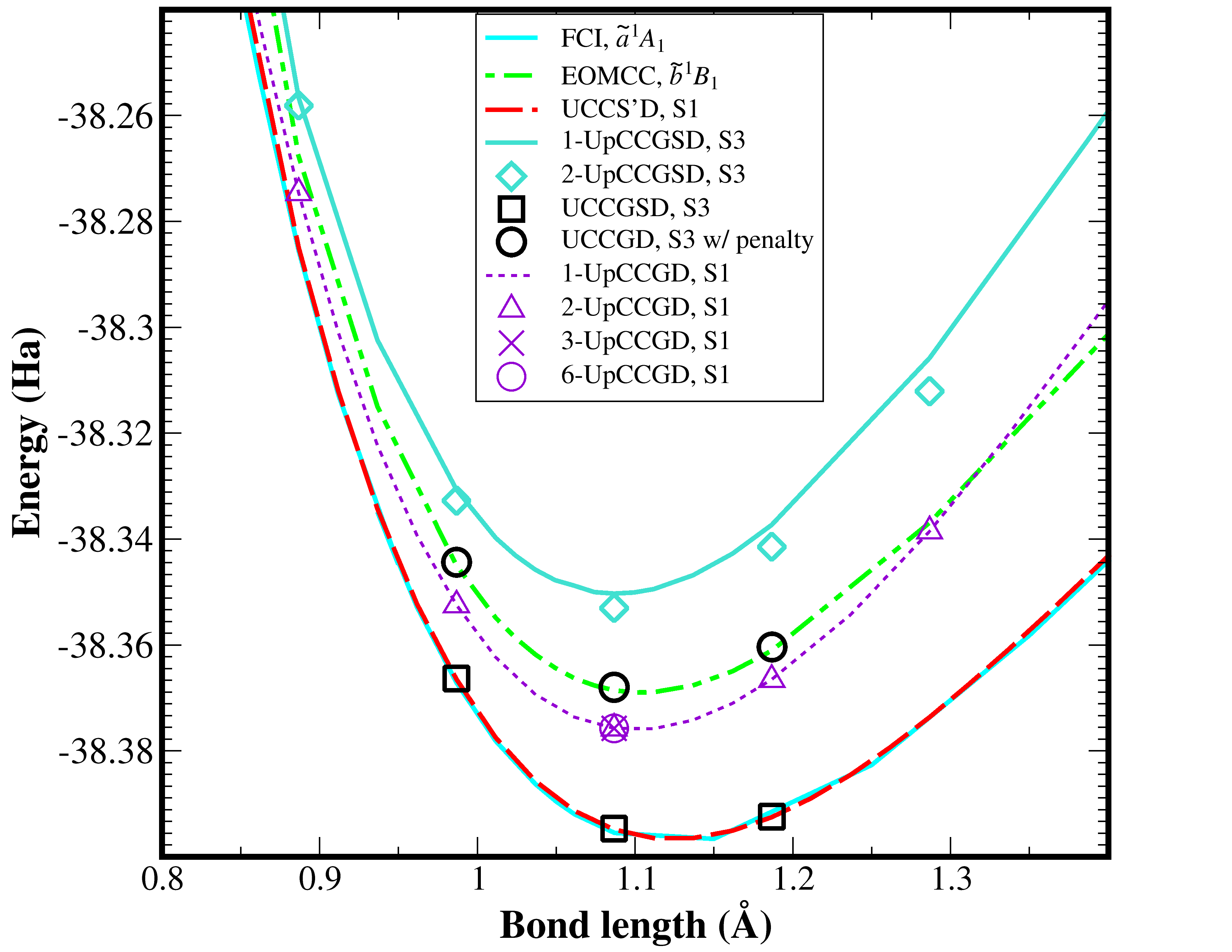}
\caption{Energy versus C-H bond length (PES) curves for the excited singlet states of CH$_{2}$ using the S1 and S3 qubit registers, calculated using various representations of the cluster operator. For \textit{k}-UpCCGSD applied to S3, the solid turquoise line denotes \textit{k} = 1 while the turquoise $\diamond$ symbols refer to results for \textit{k} = 2. Violet dashed line and symbols refer to the generalized pair-doubles only ansatz \textit{k}-UpCCGD applied to S1 (similar results are also observed for \textit{k}-UpCCGD applied to S2). The open shell singlet is obtained by the UCCGD ansatz (black $\bigcirc$) applied to the S3 register, while UCCGSD (black $\square$) falls to the closed shell singlet. Singlet classical curves (FCI and EOMCC) and results using the S1 register are also included for comparison.}
\label{ch2_curves_singlets}
\end{figure}

Two more variations of the generalized ans\"{a}tze were also considered. First, the $k$-UpCCGD operator, where singles have been removed. In this case we again see no spin crossover to the triplet ground state, confirming the major role that single excitations play in spin crossover. However, as observed for $k$-UpCCS$^{\prime}$GD, we again see that convergence with respect to $k$ is significantly affected (see Figure \ref{ch2_curves_singlets}). However, if the UCCGD operator is used, we observe robustness to spin crossover and convergence towards FCI accuracy when applied to closed shell singlet states, giving essentially the same accuracy as UCCGSD for the S2 case (see Figure \ref{ch2_curves_BL-only}). This indicates the importance of single excitations in spin cross over during VQE optimization, and their important role in \textit{k}-convergence in the \textit{k}-UpCC ans\"atze. Further insight into these findings is presented in section \ref{discussion}. 

\begin{table}
\centering
\caption{Qubit registers encoding the initial ansatz, and the resulting symmetry of the state following VQE optimization using various unitary cluster operators, without the application of spin symmetry constraints. No UCCSD results for registers S2 and S3 are included, as this operator is not suited for multireference states. The $k$-UpCCGS$^{\prime}$D and $k$-UpCCGD are not applied to the open shell singlet qubit register S3 as these ans\"{a}tze contain only those excitations corresponding to closed shell singlet configurations. The ansatz $k$-UpCCSGD, corresponding to non-generalized singles, is not applied to any multireference configuration. A mixing of the singlet and triplet states ($\tilde{a}^1A_1$ / $\tilde{X}^3B_1$) is observed when the either 1-UpCCGSD or 1-UpCCSGD is applied to the S1 qubit register, indicating spin contamination in these cases.}
\begin{tabular}{cccccc} 
\hline
\textit{initial qubit register} & UCCSD & 1-UpCCGSD & 2-UpCCGSD & 3-UpCCGSD & UCCGSD \\
\hline
\midrule
S1  & $\tilde{X}^3B_1$ & $\tilde{a}^1A_1$ / $\tilde{X}^3B_1$ & $\tilde{X}^3B_1$ & $\tilde{X}^3B_1$ & $\tilde{X}^3B_1$ \\ [5pt]
S2  & & $\tilde{a}^1A_1$ & $\tilde{X}^3B_1$ & $\tilde{X}^3B_1$ & $\tilde{a}^1A_1$ \\ [5pt]
S3  & & $\tilde{b}^1B_1$ & $\tilde{b}^1B_1$ & $\tilde{a}^1A_1$ & $\tilde{a}^1A_1$ \\ [10pt]
\hline
\textit{initial qubit register} & & 1-UpCCSGD & 2-UpCCSGD & 3-UpCCSGD &  \\
\hline
\midrule
S1  & & $\tilde{a}^1A_1$ / $\tilde{X}^3B_1$ & $\tilde{X}^3B_1$ & $\tilde{X}^3B_1$ & \\ [10pt]
\hline
\textit{initial qubit register} & UCCS$^{\prime}$D & 1-UpCCGS$^{\prime}$D & 2-UpCCGS$^{\prime}$D & 3-UpCCGS$^{\prime}$D & \\
\hline
\midrule
S1  & $\tilde{a}^1A_1$ & $\tilde{a}^1A_1$ & $\tilde{a}^1A_1$ & $\tilde{a}^1A_1$ & \\ [5pt]
S2  & & $\tilde{a}^1A_1$ & $\tilde{a}^1A_1$ & $\tilde{a}^1A_1$ & \\ [10pt]
\hline
\textit{initial qubit register} & & 1-UpCCGD & 2-UpCCGD & 3-UpCCGD & UCCGD \\
\hline
\midrule
S1  & & $\tilde{a}^1A_1$ & $\tilde{a}^1A_1$ & $\tilde{a}^1A_1$ & $\tilde{a}^1A_1$ \\ [5pt]
S2  & & $\tilde{a}^1A_1$ & $\tilde{a}^1A_1$ & $\tilde{a}^1A_1$ & $\tilde{a}^1A_1$ \\ [5pt]
S3  & & & & & $\tilde{X}^3B_1$ \\ [5pt]
\end{tabular}
\label{spin_cross}
\end{table}

\begin{figure}[ht]
\centering
\includegraphics[width=14cm]{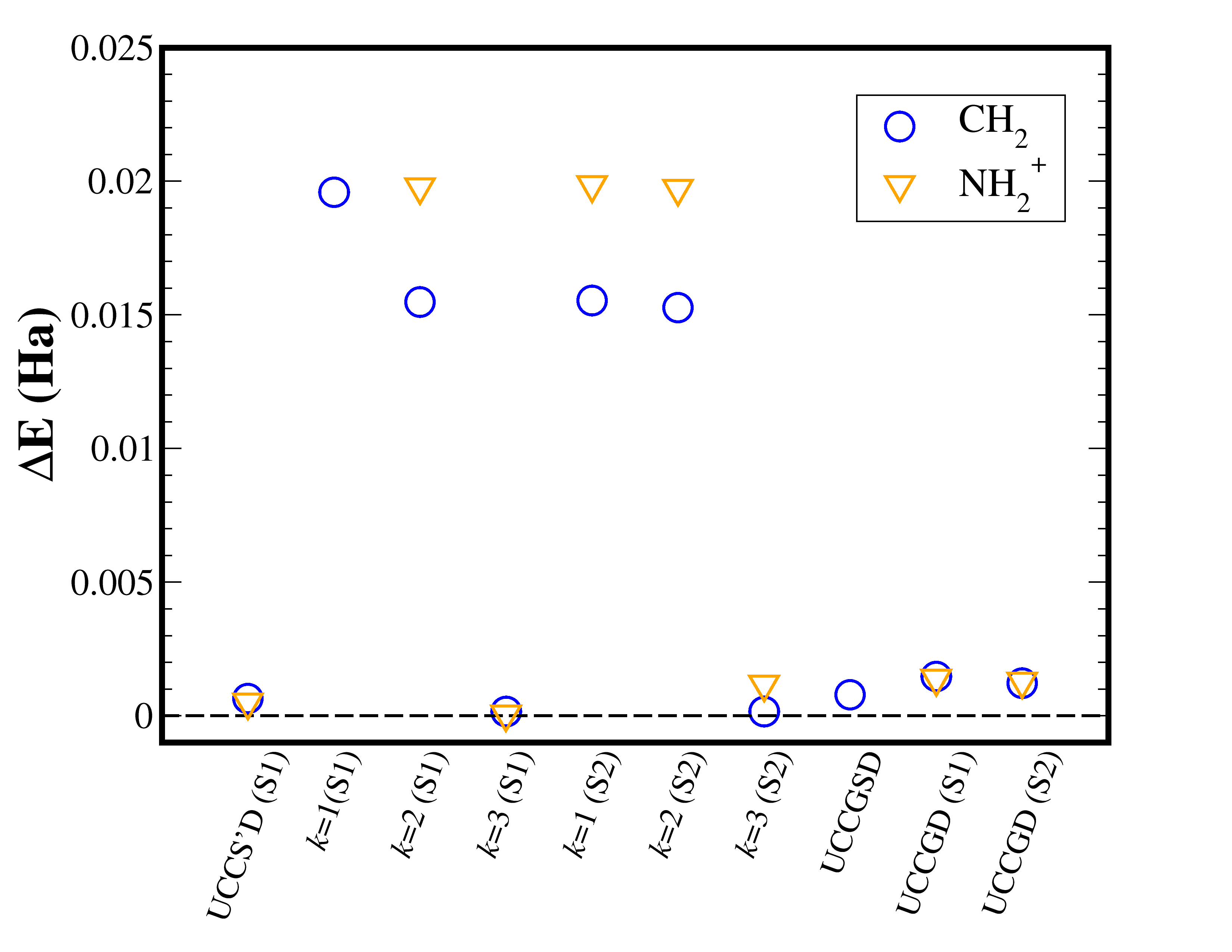}
\caption{Calculated energy of the $\tilde{a}^1A_1$ singlet state for CH$_2$ (blue $\bigcirc$ symbols) and NH$_{2}^{+}$ (orange $\triangledown$ symbols) relative to the corresponding FCI energy at the minimum energy bond length, for various UCC ans\"atze optimized with the VQE. Values of \textit{k} shown on the horizontal axis refer to \textit{k}-UpCCGS$^{\prime}$D calculations. The singlet spin symmetry is maintained for all calculations shown here.}
\label{ch2_nh2plus_fcicomparison}
\end{figure}

We now address the accuracy of these methods with respect to the FCI result. For the \textit{k}-UpCCGS$^{\prime}$D energy curves, the value of \textit{k} has a lower effect on the optimized energies throughout the PES than the $\tilde{X}^3B_1$ ground state. Compared to UCCGSD (S2) or UCCGD (S1 and S2), \textit{k}-UpCCGS$^{\prime}$D performs slightly worse for the excited singlet states. At the PES minimum, the singlet calculated with the S1 register is 0.02 Ha above the FCI limit for \textit{k} = 1, while this difference decreases to 0.016 Ha for \textit{k} = 2. We observe a similar behaviour for the curve calculated with the S2 register within \textit{k}-UpCCGS$^{\prime}$D, however \textit{k} = 1 and \textit{k} = 2 yield the same PES minimum energy to within 1 mHa. We attribute the relatively poor accuracy of \textit{k}-UpCCGS$^{\prime}$D to the presence of two molecular bonds stretching simultaneously: this is one of the situations in which this ansatz is supposed to struggle to find accurate results \cite{lee_kupcc}. Convergence towards the FCI limit is reached for $k$ = 3 , with a difference of less than 1 mHa for either the S1 or S2 qubit registers. For UCCGSD or UCCGD with the qubit registers T, S1, and S2, the ground state triplet and excited state singlet are accurately calculated; this shows that fully generalized double excitations are adequate to capture the ground and first excited state of this system. Figure \ref{ch2_nh2plus_fcicomparison} presents the calculated $\tilde{a}^1A_1$ singlet energies for CH$_{2}$ relative to FCI for the various UCC ans\"atze.

 We proceed now to study qubit register S3, which should represent the energy surface of the open shell multireference singlet $\tilde{b}^1B_1$. The initial qubit register was prepared using the quantum circuit displayed on the bottom panel of Figure \ref{lc-csS_circuit}, and the cluster operators 1-UpCCGSD, 2-UpCCGSD, 3-UpCCGSD, UCCGSD, and UCCGD were applied to it. For 2-UpCCGSD, five points around the minimum of the PES were calculated, while for UCCGSD only 3 points were calculated due to computational expense. These results were compared with the $\tilde{b}^1B_1$ curve obtained by classical EOMCC calculations. The 1-UpCCGSD and 2-UpCCGSD ans\"{a}tze manage to conserve the $\tilde{b}^1B_1$ character of the wavefunction, but tend to overestimate the energy of this state relative to EOMCC, as shown in Figure \ref{ch2_curves_singlets}. Switching to 3-UpCCGSD leads to the loss of the $\tilde{b}^1B_1$ symmetry and convergence to the $\tilde{a}^1A_1$ state. This result suggests that the retention of $\tilde{b}^1B_1$ for 1-UpCCGSD and 2-UpCCGSD is likely due to a local minimum in the energy optimization resulting from insufficient $k$ factors to achieve full variational freedom. This is also consistent with the observation of lower accuracy for the $\tilde{b}^1B_1$ energies from 1-UpCCGSD and 2-UpCCGSD shown in Figure \ref{ch2_curves_singlets}. Accordingly, we find that the UCCGSD ansatz calculation converges to the closed shell singlet $\tilde{a}^1A_1$, which is not unexpected as the fully generalized UCCGSD operator should indeed converge to the variational minimum for a given symmetry. No spin crossover to the triplet state is observed for any of these ans\"{a}tze with this qubit register. This is at variance to the S1 or S2 cases. However, when UCCGD is applied to the S3 register, the combination of single occupied spatial orbitals and non-paired double excitations drives the VQE optimization towards the low spin S$_z$ = 0 component of the triplet ground state. This crossover is only prevented by applying a spin penalty term in the VQE optimization; in that case, the energy curve for the UCCGD ansatz converges to the EOM-CC curve. Further studies are required to understand these aspects of the various cluster operators and qubit register representations of the singlets, and this will be explored in a future work.

To conclude this section, our results show that generalized cluster operators lead to a strong dependence of the performance of the VQE algorithm (i.e. whether the algorithm can reach the desired solution) on the symmetry properties of the initialized state (single reference or multireference, closed shell or open shell). In general, for the VQE algorithm both the initial wavefunction ansatz $\ket{\Psi_{0}} = U_{0}\ket{\text{HF}}$ and the orbitals that enter the one- and two-electron (Hartree-Fock) integrals for the second quantization Hamiltonian $h_{mol}$ (given as input to the optimization routine) usually share the same spin symmetry. Here, $U_{0}$ represents the unitary cluster operator in which the excitation amplitudes have their initially randomized values. To further investigate the dependence of the VQE algorithm on the initial state, a test was carried out to qualitatively check the relative contributions of $\ket{\Psi_{0}}$ and $h_{mol}$ to the VQE-calculated total energy. To this end, $\ket{\Psi_{0}}$ was initialized to a closed shell singlet (via circuit preparation for $\ket{\text{HF}}$), while the integrals in $h_{mol}$ were obtained from a Hartree-Fock calculation with the molecular orbitals arranged in a spin triplet. This test was performed for a range of C-H bond lengths around the PES minimum using 1-UpCCGSD. The optimized energies from these calculations correspond very closely (to less than 1 mHa) to the singlet excited state for the same cluster operator (results are not plotted in Figure \ref{ch2_curves_BL-only} for clarity). Moreover, the initial (unoptimized) energy for this calculation is only 3 mHa above that of the initial energy when the singlet configuration is used for $\ket{\Psi_{0}}$ and $h_{mol}$. This strongly indicates that the initial wavefunction ansatz is by far the stronger contributor to the total energy calculated from the VQE procedure.

\subsubsection{Charged states calculation}

\noindent \textit{k}-UpCCGSD calculations were also performed for the anionic (CH$_{2}^-$) and cationic (CH$_{2}^+$) states of CH$_{2}$. At the minimum of the PES curves the CH$_{2}^-$ and CH$_{2}^+$ ground states correspond to spin doublets with energies of -38.1498 Ha and -38.1453 Ha, respectively for \textit{k} = 1. For \textit{k} = 2, the minimum total energy for CH$_{2}^-$ (CH$_{2}^+$) is lowered by 0.005 Ha (0.008 Ha) relative to \textit{k} = 1. When UCCSD is adopted, the FCI limit for the charged states is recovered. It is interesting to note that the ionic states remain separated from the neutral single reference and multireference singlet states by a large amount throughout the PES curve; at the minima both ionic states have energies approximately 0.22-0.23 Ha above that of the single reference singlet, when comparing results from UCCSD calculations. Thus, these calculations show that the neutral singlet state is not accessed by either charged ionic doublet via unconstrained VQE optimization. 

\begin{figure}[ht]
\centering
\includegraphics[width=14cm]{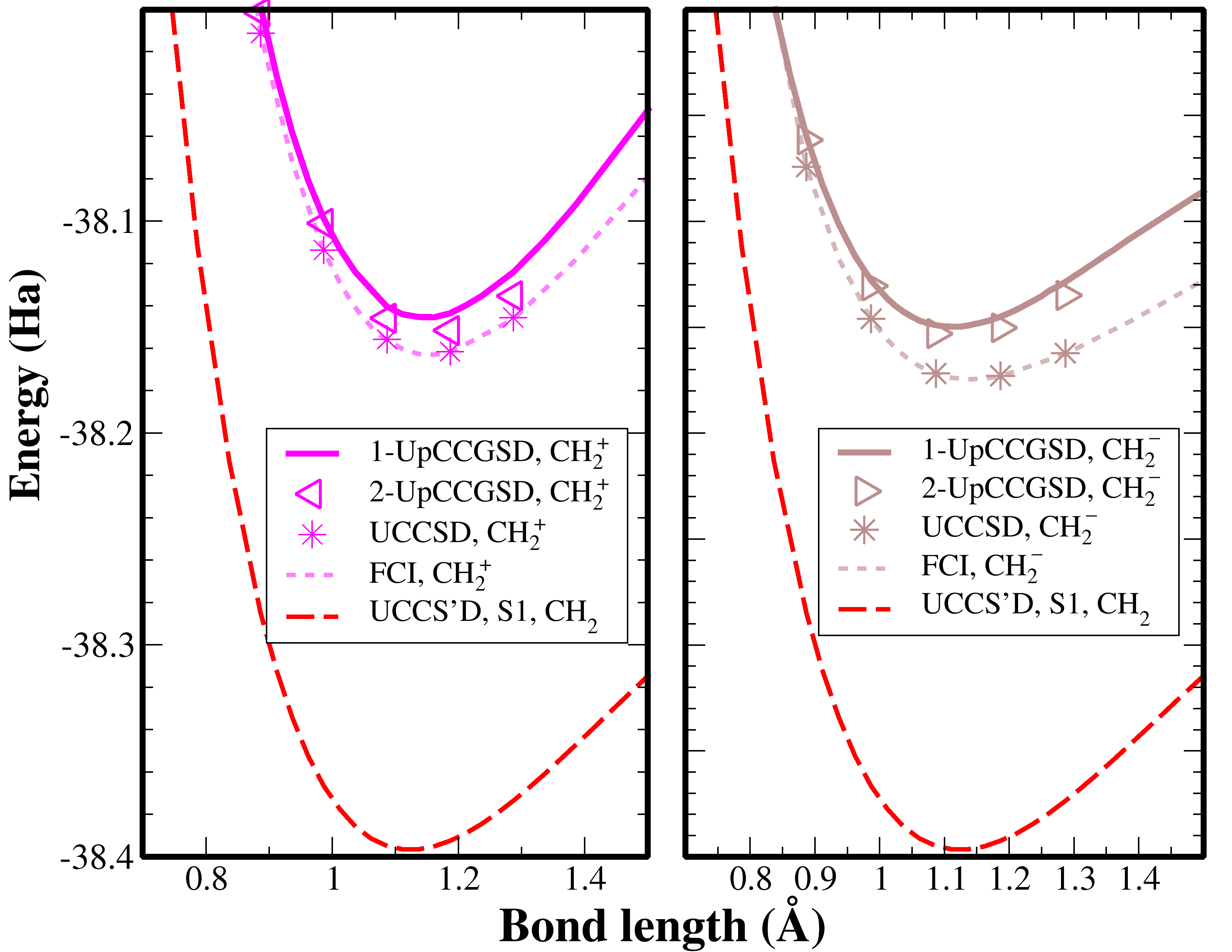}
\caption{Energy versus C-H bond length (PES) curves for charged molecules CH$_{2}^+$ (left panel) and CH$_{2}^-$ (right panel) calculated using the UCCSD and 1-UpCCSD cluster operators. These are also compared to classical calculations and the UCCSD calculation of the $\tilde{a}^1A_1$ neutral state.}
\label{ch2_curves_ions}
\end{figure}

\begin{figure}[ht]
\centering
\includegraphics[width=14cm]{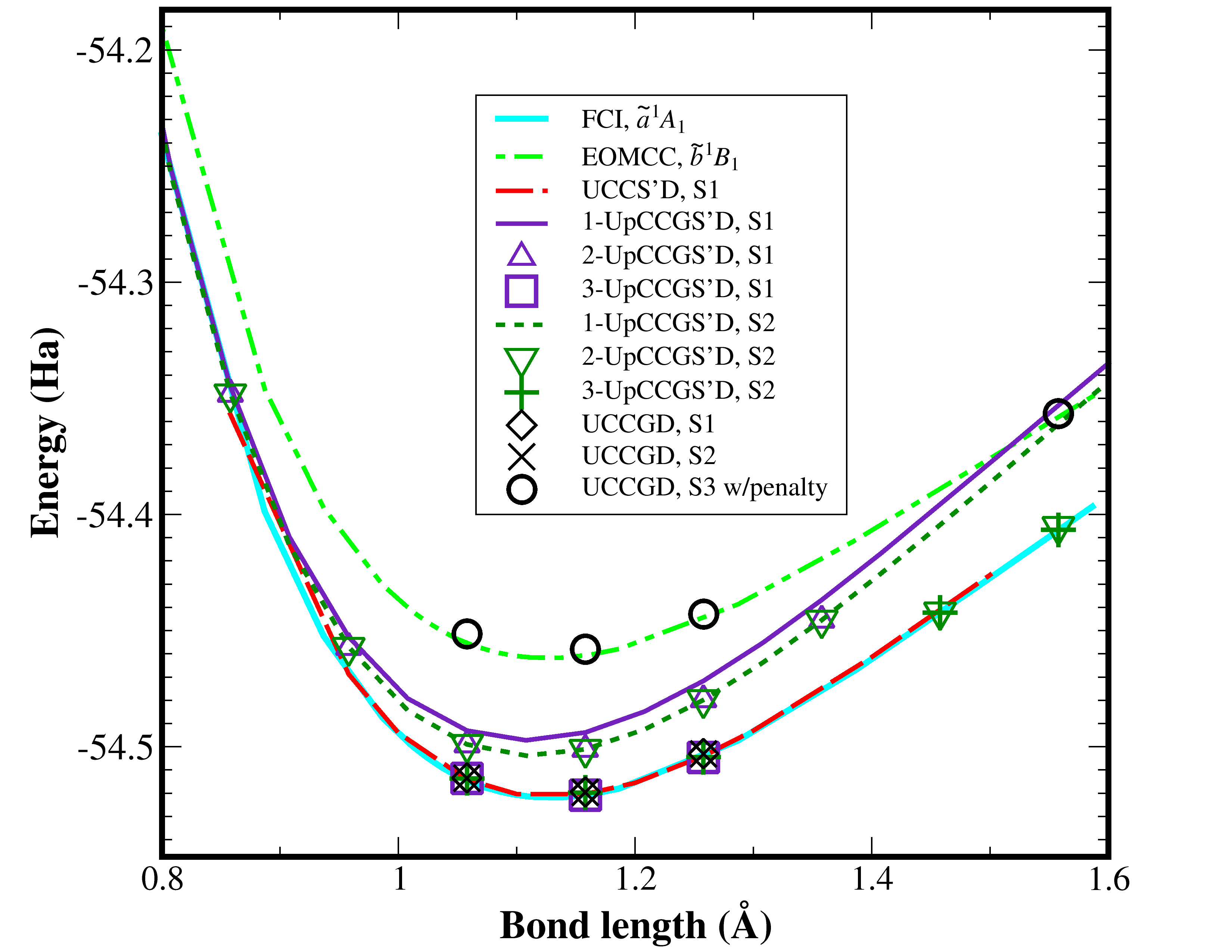}
\caption{Energy versus N-H bond length (PES) curves for the ground and excited states of NH$^{+}_{2}$, calculated using various representations of the cluster operator, and using different initial configurations of spin orbitals as input to VQE optimization. For \textit{k}-UpCCGS$^{\prime}$D, lines denote results for \textit{k} = 1, while the symbols of the corresponding color refer to results for \textit{k} = 2 or fully generalized ans\"{a}tze. The green $+$ and cyan $\times$ symbols refer to the initial state configurations S2 and S3 respectively, calculated using UCCGD.
}
\label{nh2plus_curves_BL-only}
\end{figure}

\begin{figure}[ht]
\centering
\includegraphics[width=14cm]{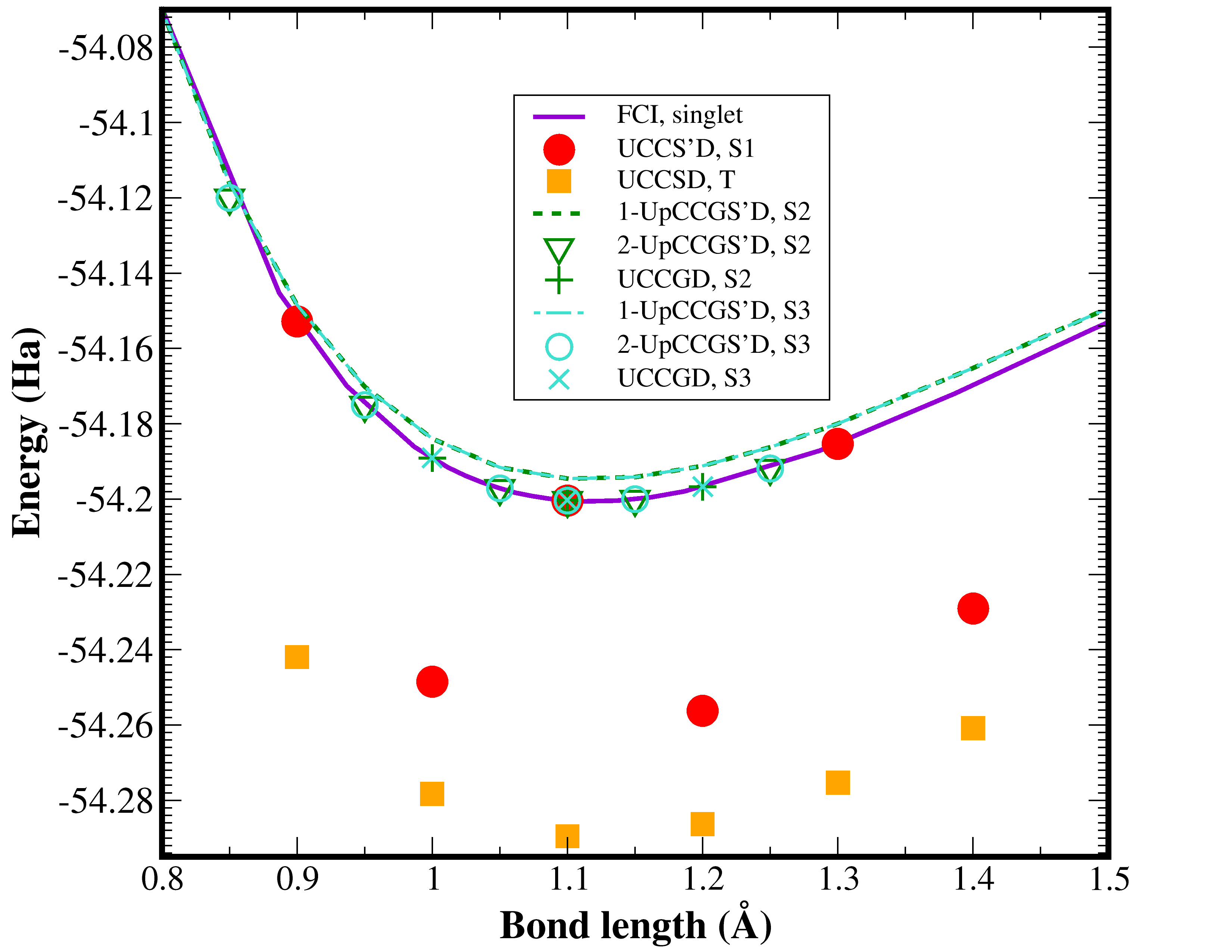}
\caption{Energy versus N-H bond length (PES) curves for the ground and excited states of NH, calculated using various representations of the cluster operator, and using different initial configurations of spin orbitals as input to VQE optimization. For \textit{k}-UpCCGS$^{\prime}$D, lines denote results for \textit{k} = 1, while the symbols of the corresponding color refer to results for \textit{k} = 2 or fully generalized ans\"{a}tze. The green $+$ and cyan $\times$ symbols refer to the initial state configurations S2 and S3 respectively, calculated using UCCGD.
}
\label{nh_curves_BL-only}
\end{figure}
 
\begin{figure}
\centering
\includegraphics[width=14cm]{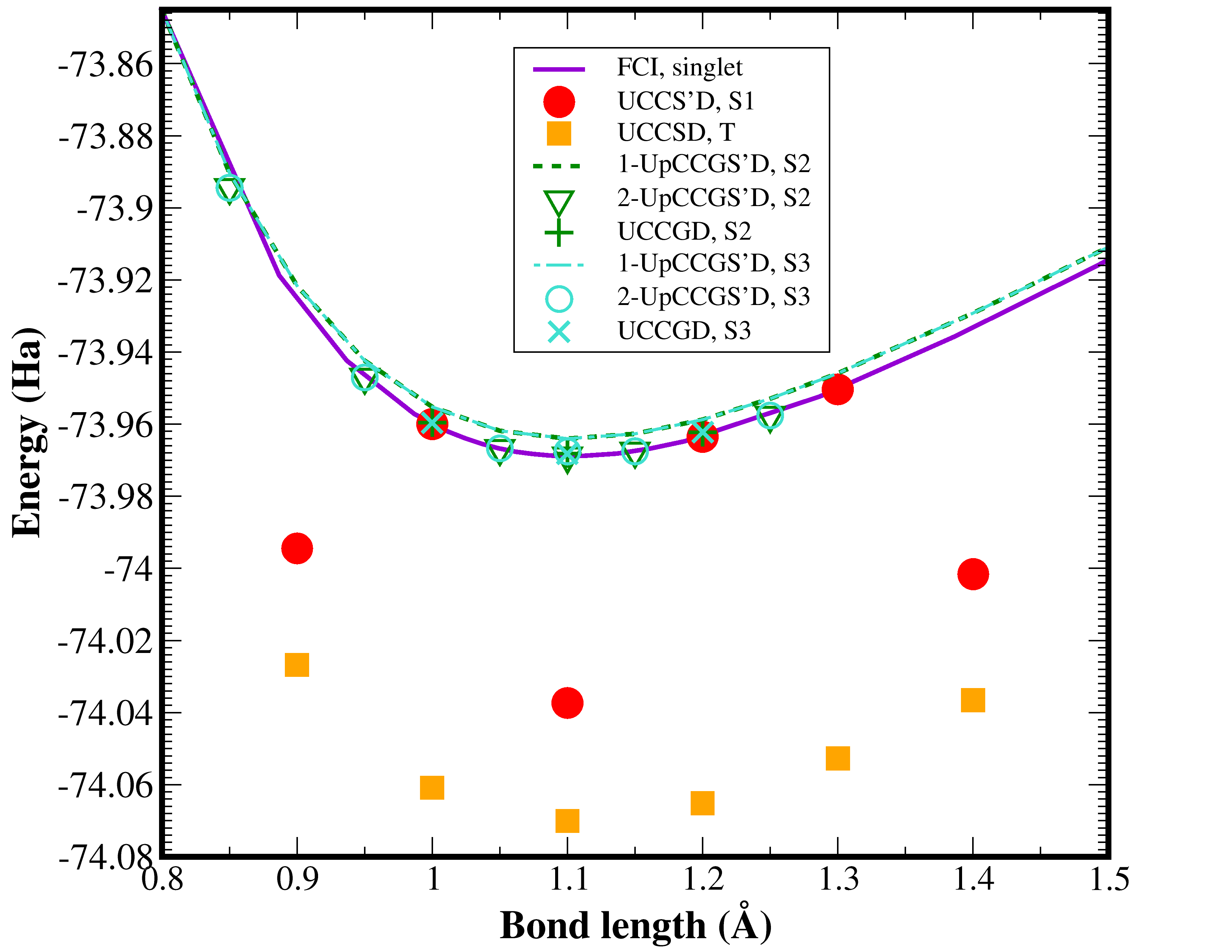}
\caption{Energy versus O-H bond length (PES) curves for the ground and excited states of OH$^+$, calculated using various representations of the cluster operator, and using different initial configurations of spin orbitals as input to VQE optimization. For \textit{k}-UpCCGS$^{\prime}$D, lines denote results for \textit{k} = 1, while the symbols of the corresponding color refer to results for \textit{k} = 2 or fully generalized ans\"{a}tze. The green $+$ and cyan $\times$ symbols refer to the initial state configurations S2 and S3 respectively, calculated using UCCGD.
}
\label{ohplus_curves_BL-only}
\end{figure}

\begin{figure}
\centering
\includegraphics[width=14cm]{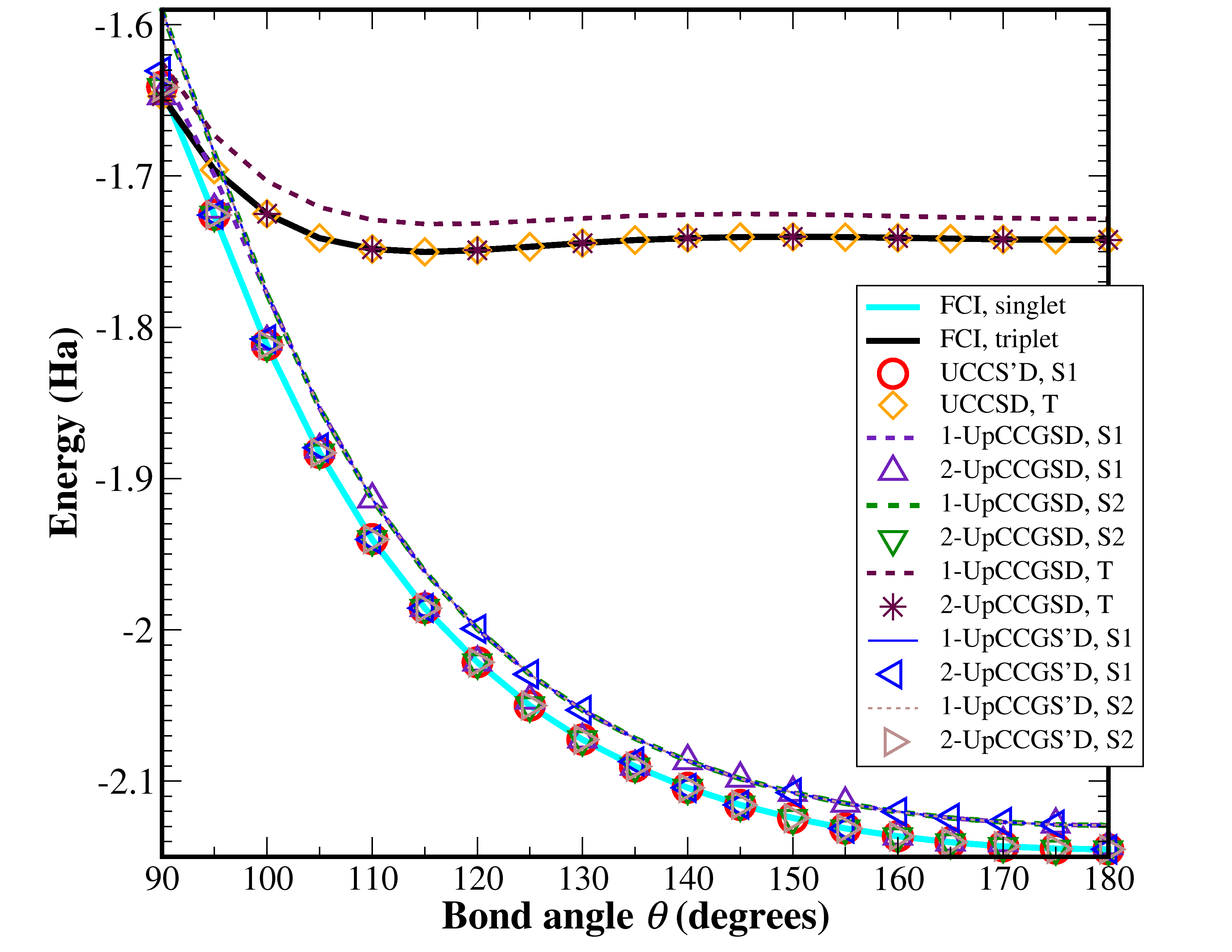}
\centering
\includegraphics[width=10cm]{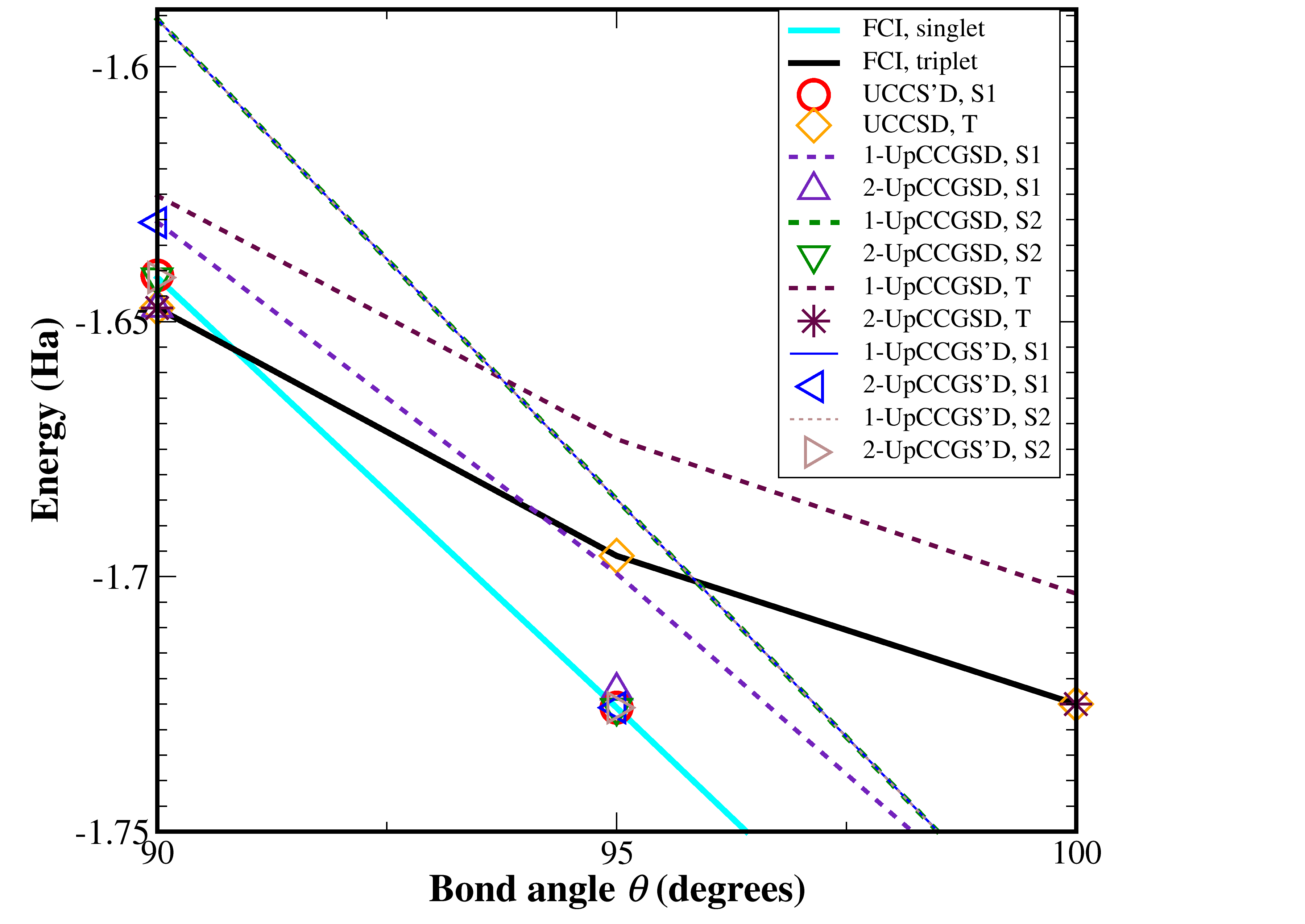}
\caption{Top panel shows energy versus bond angle for the ground and excited states of H$_4$, calculated using various representations of the cluster operator, and using different initial configurations of spin orbitals as input to VQE optimization. For \textit{k}-UpCCGSD and \textit{k}-UpCCGS$^{\prime}$D, lines denote results for \textit{k} = 1, while the symbols of the corresponding color refer to results for \textit{k} = 2. Bottom panel shows the same results but for H-H-H bond angles close to 90$^{\circ}$.
}
\label{h4_curves_BL-only}
\end{figure} 

\subsection{NH$^{+}_{2}$}

Next, the diradical NH$^{+}_{2}$ is studied, focusing on its low lying excited singlet states to investigate the performance of ans\"{a}tze that were observed to be robust to spin cross over for CH$_2$. Thus, UCCS$^{\prime}$D, $k$-UpCCGS$^{\prime}$D, and UCCGD are used for the close shell singlet registers S1 and S2, while UCCGD with a spin penalty is used for the open shell singlet register S3. As the excited singlet states are investigated here, the N-H-N angle is fixed to 107$^{\circ}$, corresponding to the minimum energy structure for the singlet \cite{slipchenko_ch2}. As in the previous section, a spin penalty to calculate the energy as a function of bond length is used for the generalized doubles only UCCGD applied to the qubit registers S3, while for the T, S1 and S2 curves presented in Figure \ref{nh2plus_curves_BL-only} no spin penalty term is applied. 

Similar results are observed here as for CH$_2$. The UCCS$^{\prime}$D operator results in energies reasonably close to FCI for the closed shell singlet, and the $k$-UpCCGS$^{\prime}$D ansatz achieves roughly 1 mHa or less above FCI for $k$ = 3 using either the S1 or S3 qubit registers. Similar accuracy is observed for UCCGD applied to S1 and S2. While for UCCGD applied to S3, the EOMCC result is approached when a spin penalty is applied to the VQE optimization. Figure \ref{ch2_nh2plus_fcicomparison} presents the calculated $\tilde{a}^1A_1$ singlet energies for NH$_{2}^{+}$ relative to FCI for the various UCC ans\"atze. For the $k$-UpCCGS$^{\prime}$D ansatz, it is observed that at \textit{k} = 1, the S1 and S2 curves cross the $\tilde{b}^1B_1$ state at N-H bond lengths of approximately 1.5 \AA. To investigate this, larger bond lengths were calculated for \textit{k} = 2 and 3 using the S2 register, and in these cases the calculated energy falls to the within 2 (1) mHa of the FCI values for \textit{k} = 2 (3). Thus (as also observed below for the H$_4$ case, shown in Figure \ref{h4_curves_BL-only}) the convergence with respect to \textit{k} can occasionally be dependent on the geometry, and the crossing of the \textit{k} = 1 S1 and S2 curves with the $\tilde{b}^1B_1$ state is an artifact of the poor description of this state at lower values of \textit{k}. Interestingly, the UCCGD ansatz constrained with a spin penalty maintains $\tilde{b}^1B_1$ symmetry (unlike $k$-UpCCGS$^{\prime}$D with spin constrained singles, or UCCGSD with a spin penalty) at bond lengths greater than 1.5 \AA.

\subsection{NH and OH$^{+}$}

As in the previous section, here only the results for UCCS$^{\prime}$D, $k$-UpCCGS$^{\prime}$D, and UCCGD are shown and compared to the FCI limit for the excited singlet state. The results here again indicate that the FCI limit of the excited state can be reached, without the need of spin penalty functions to constrain the $S^2$ operator, by restricting excitations in the cluster operator and applying regular VQE optimization with the appropriate qubit register to encode the initial state. Unlike the previous examples, for these linear diradical molecules NH and OH$^+$, the first excited state is an open shell diradical singlet \cite{slipchenko_ch2, ess11, ess12}, and in our calculations the first excited state is accessed with good accuracy relative to FCI for all ans\"atze apart from UCCS$^{\prime}$D. 

The failure of UCCS$^{\prime}$D to capture the first excited state with a single reference state is particularly interesting since this operator is also implemented to restrict single excitations to closed shell singlet configurations (as in equation \ref{eqn_k-up_gsprime_T} although with non-generalized excitations). For the calculations in which UCCS$^{\prime}$D does not retain the singlet symmetry, the total spin operator squared has a value $\bra{\Psi}S^2\ket{\Psi}$ $\simeq$ 1.45 following VQE optimization initialized with the S1, while for the singlet (triplet) a value of 0 (2) is expected. In these cases (red circles on Figure \ref{nh_curves_BL-only} and Figure \ref{ohplus_curves_BL-only}) the dominant determinants after optimization correspond to the $S_z = 0$ component of the triplet, along with smaller but significant contributions from closed shell singlets, indicating a breakdown of the desired singlet symmetry in favor of a mixture of singlet and triplet symmetries. Similar findings are observed for OH$^+$ in those cases when the singlet symmetry is lost due to heavy spin contamination.

In both cases of NH and OH$^+$, applying $k$-UpCCGS$^{\prime}$D to the S1 register (not shown for clarity) does not lead to a loss of singlet symmetry following VQE optimization, but in fact retains $\bra{\Psi}S^2\ket{\Psi}$ = 0 and reproduces the energies obtained for S2 and S3 qubit registers. In addition, the qubit registers corresponding to multireference states (which are not compatible with UCCSD) are also robust to this error, and converge to the FCI limit of the first excited state using either the closed shell S2 or open shell S3 configurations to initialize the VQE routine, hence single reference or multireference initial states can both access the first excited state once generalized excitations are used. In addition, all UCCGD calculations are also robust to the loss of singlet spin symmetry; comparing this finding to the UCCS$^{\prime}$D and $k$-UpCCGS$^{\prime}$D results shows that the combined use of singles and unpaired double excitations can lead to the loss of singlet spin symmetry when the excited singlet exhibits an open shell configuration.

In summary, the first excited state of NH and OH$^+$ can be accessed with good accuracy relative to FCI once the excitations are restricted in the following manner: if single excitations are present then the singles must be restricted as in equation \ref{eqn_k-up_gsprime_T} and doubles must be restricted to paired doubles, otherwise if all single excitations are omitted then generalized doubles can access the diradical singlet state. Taking these considerations into account, it is clear that the UCCS$^{\prime}$D operator can fail to access the first excited state, at variance to the diradical CH$_2$ and NH$^{+}_{2}$ cases for which the first excited state has a closed shell symmetry \cite{slipchenko_ch2}. This is exemplified by the results presented in this section and also discussed in more detail in section \ref{discussion}.
 
\subsection{H$_4$}

With the H-H distance fixed to 0.75 \AA, H$_4$ in a square geometry has a triplet ground state, although the triplet energy is quite close to that of the singlet (the latter is approximately 7 mHa above the triplet for the square geometry). As the H-H-H bond angle $\theta$ is increased from 90$^{\circ}$ (square) to $>$ 90$^{\circ}$ (trapezoid) the singlet state quickly becomes more stable, and at linear geometry (bond angle = 180$^{\circ}$), the ground state singlet is well separated from the triplet (see Figure \ref{h4_curves_BL-only}). We use the energy dependence of the bond angle to compare the standard UCCSD operator,  $k$-UpCCGSD, and the modified version $k$-UpCCGS$^{\prime}$D investigated in previous sections. 

The key results we obtain for the H$_4$ system are as follows. The UCCSD operators reproduce the FCI energy curves for the singlet and triplet states. For the $k$-UpCCGSD ansatz applied to the single reference register S1, the effects of spin contamination are observed for 90$^{\circ}$ $\leq$ $\theta$ $<$ 100$^{\circ}$. At $\theta$ = 90$^{\circ}$ the spin operator value $\bra{\Psi}S^2\ket{\Psi}$ is approximately 1 following VQE optimization of the S1 register using $k$-UpCCGSD. This is observed for 1-UpCCGSD and 2-UpCCGSD applied to the S1 register, and is further exemplified by the close correspondence of energy of these curves (purple dashed line and triangle) with the energy curves of the $k$-UpCCGSD ans\"{a}tze (maroon dashed line and star symbol) applied to the T register (triplet) for small angles, see Figure \ref{h4_curves_BL-only}. As $\theta$ increases to the value at which the singlet is stable, $\bra{\Psi}S^2\ket{\Psi}$ = 0 and the singlet symmetry is recovered for the $k$-UpCCGSD applied to the S1 register. For the multireference S2 register, the singlet symmetry is maintained ($\bra{\Psi}S^2\ket{\Psi}$ = 0) for $k$ = 1 and $k$ = 2, and the singlet energy is above that of the triplet for the same UCC ansatz. Again, we observe robustness to spin crossover/contamination when the modified ansatz $k$-UpCCGS$^{\prime}$D is used, for all values of $k$, such that the triplet is more stable than the singlet for the single reference initial state S1 as well as for S2, when single excitations are restricted as in equation \ref{eqn_k-up_gsprime_T}.  

For either $k$-UpCCGSD or $k$-UpCCGS$^{\prime}$D ans\"{a}tze we observe incorrect or poor convergence, respectively, of the energy curve with respect to $k$ when the single reference configuration S1 is used to represent the singlet. For the multireference configuration S2, both ans\"{a}tze yield good accuracy with respect to FCI for $k$ = 2. This is a further indication of the interesting behaviour that occurs when spin restricted generalized ans\"{a}tze are applied to multireference wavefunctions; the restriction of single excitations protects against spin cross over to the triplet of the excited singlet state (where unrestricted single excitations can change the spin symmetry), while initializing the optimizer with a multireference state increases the accuracy of the converged result. 

\begin{figure}[ht]
\centering
\includegraphics[width=14cm]{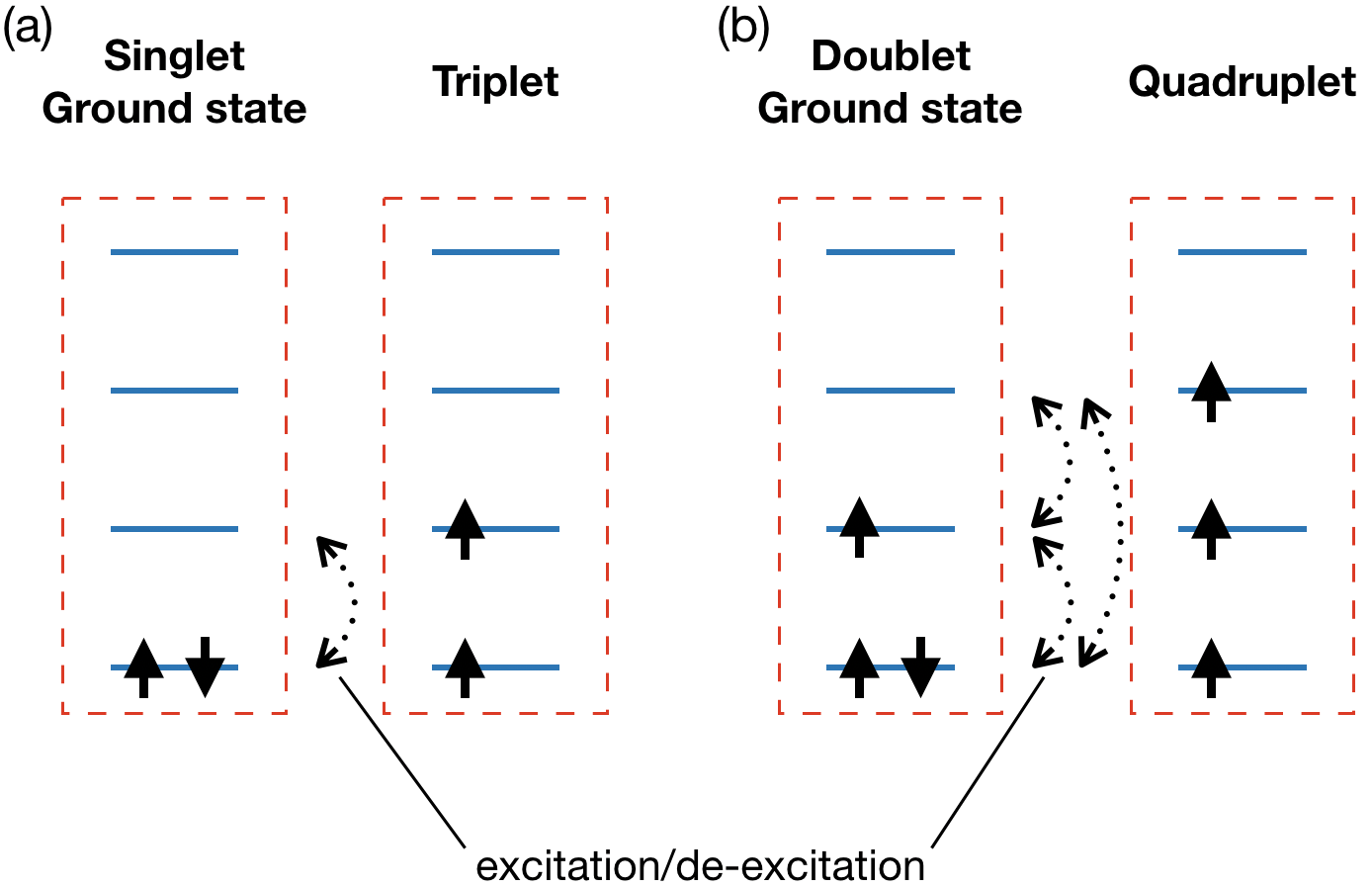}
\caption{Spin configurations for the 4 lowest spatial orbitals of the ground and first excited states of H$_{2}$ (a) and H$_{3}$ (b), showing the excitations or de-excitations necessary to reach one state from the other. For both molecules, at least one spin flipping (de-)excitation is needed for the ground-excited state transition.}
\label{h2_h3_spin-configs}
\end{figure}

\section{Discussion}\label{discussion}

In general, our results for the molecules considered in this study show that VQE is capable of calculating excited and ionic states, provided their spin or particle number symmetry is different from that of the ground state. In this sense, it is instructive to compare the results obtained with different ans\"{a}tze of fermionic character, like the ones analyzed in this work, with the results obtained with hardware-efficient ans\"{a}tze, where the physical character of the resulting parametrized circuit is more difficult to evaluate \cite{kandala_ansatz}. It has been demonstrated that this last type of ansatz may suffer from convergence to unwanted states due to a lack of constraints which cause VQE optimization to tend toward energy states of different symmetry than the ground state of interest, as these states may be lower in energy at certain regions of the PES \cite{ryabinkin_penalties}. Our results for different types of UCC-inspired operators acting on a Hartree-Fock wavefunction indicate that the particle number is robust against this problem when using ans\"{a}tze based on physical principles. This stems from the fact that all entangling operators considered commute with the number operator. If the number of electrons is well defined, as a result of encoding the Hartree-Fock wavefunction in the initial qubit register, then the cluster operator will conserve particle number when calculating the entangled wavefunction. However, if one allows the preparation of the mean-field (e.g. Hartree-Fock) state to be variationally optimized (using parameterized R$_{X}(\theta)$ rotation gates), then switching to states with different electron numbers becomes possible. In fact, we can reproduce the transition from the neutral triplet to the cation doublet reported at Ref \cite{ryabinkin_penalties}, and observe the "kink" at bond length $\sim$0.7 \AA, for H$_{2}$ using this method. This transition does not occur, and no switching between states is observed for the molecules considered when the Hartree-Fock state is prepared by non-parameterized gates which flip the desired number of qubits to correspond to the desired number of occupied spin orbitals (e.g. X gates).

Our calculations for diradical species show, however, that the UCC-inspired ans\"{a}tze may require additional constraints to obtain the desired spin symmetry in the PES. It is not obvious at first glance whether these ans\"{a}tze will yield the correct energy surface for a given set of symmetries, and this constitutes a strong argument to combine them with symmetry-constraining methods to ensure that the correct results are achieved. Clearly, the required symmetry constraint depends on the system; for example, when the desired excited state has a (single or multireference) closed shell singlet character, restricting the set of single excitations is sufficient to prevent spin crossover to a different spin symmetry (e.g. triplet) and achieve the FCI limit for the singlet, without the need for symmetry constraining penalty terms in the VQE optimization. It has long been recognised that knowledge of the character of the desired excited state can be utilized to enhance the efficiency of classical computations. Our results indicate that knowledge of the excited state spin configuration can also significantly enhance the efficiency of VQE optimization, as well as allow for reliable predictions of when to expect undesired spin cross over during VQE optimization of the cluster amplitudes and which states will be protected from spin cross over even without spin penalties or excitation constraints as in equation \ref{eqn_k-up_gsprime_T}. Exclusion of excitations which cause spin cross over allow for access to the desired state without a significant loss of accuracy, in addition to a gain in efficiency due to the lower number of variational parameters. 

Prediction of spin cross over behaviour in advance of a VQE optimization can be exemplified by the first excited states of H$_{2}$, H$_{3}$, and CH$_{2}$. The first excited state of H$_{2}$ is a triplet, with spin configuration represented by Figure \ref{h2_h3_spin-configs}. As spin flip excitations are not included in any cluster operator we consider here (which is physically reasonable in the absence of magnetic interactions or spin-orbit coupling), the Pauli exclusion principle prevents parallel spins from sharing the same spatial orbital and thus the excitation resulting in a closed shell singlet configuration is avoided. For H$_{3}$, the first excited state is a quadruplet, represented as 3 parallel unpaired spins. Again, the absence of spin flipping terms prevents the VQE optimization from reaching the lower doublet state, thereby maintaining the correct spin symmetry of the first excited state. Hence, knowledge of the spin symmetry of these states in advance of the VQE calculation can be used to predict that spin cross over will be avoided without a spin penalty or excitation constraints as in equation \ref{eqn_k-up_gsprime_T}. 

For CH$_{2}$, the spin cross over behaviour can be explained in a similar manner. The first excited state in this case is a closed shell singlet and the ground state is a triplet. If one starts from a closed-shell qubit register, like the S1 or S2 registers, single excitations (non-spin flipping) can cause a transition from a singlet to a triplet, as long as $\alpha$-$\alpha$ and $\beta$-$\beta$ transitions have independent amplitudes. As seen in the previous section, the dominant determinants correspond to the $S_{z}$ = 0 component of the triplet (in terms of spin functions, $\sim\frac{1}{2}(\alpha\beta + \beta\alpha)$), and a loss of singlet spin symmetry is observed with $\bra{\Psi}S^2\ket{\Psi}$ significantly smaller than 2. Once the single amplitudes are restricted as in equation \ref{eqn_k-up_gsprime_T}, the first excited state of CH$_{2}$, and the other diradical molecules with multireference excited singlets can be accessed by regular VQE without additional constraints. This is consistent with techniques used by Evangelista et al. \cite{evangelista19} to make the UCC operator commute with the $S^{2}$ operator, in the framework of spin-adaptation methods studied in the past for coupled cluster theory \cite{knowles_sac, neogrady_sac, szalay_sac, heckert_sac}.

CH$_2$ and the other diradical species studied in this paper are also useful to illustrate the challenges at calculating open-shell singlet states with VQE. In particular for CH$_2$, it is interesting to find that the S3 qubit register can drive the optimized wavefunction to the $\tilde{b}^1B_1$ state instead of the  $\tilde{a}^1A_1$ state, which is lower in energy. This property comes at the cost of imposing extra constraints on the parametrized wavefunction, like setting low values of $k$ using $k$-UpCCGSD or allowing all double excitations while removing the single excitations, as defined in the UCCGD ansatz. However, these approximations are associated with reduced accuracy or the need to impose penalty terms to prevent spin crossover. A different picture appears when addressing the NH and OH$^+$ diradicals. In these cases, the $\tilde{b}^1B_1$ state is the lowest energy singlet \cite{slipchenko_ch2, ess11, ess12}, and the use of the UCCS$^{\prime}$D ansatz in combination with the S1 register results in heavy spin contamination and a loss of singlet spin symmetry during VQE optimization. Due to the presence of singles and unpaired double excitations in this ansatz, a large amount of spin contamination is possible because the excitations required to yield an open shell singlet from a closed shell singlet will not necessarily restrict the signs of the determinant in the expansion of the wavefunction, so both $\frac{1}{2}(\phi_{1}\phi_{2} + \phi_{2}\phi_{1})(\alpha\beta - \beta\alpha)$ and $\frac{1}{2}(\phi_{1}\phi_{2} - \phi_{2}\phi_{1})(\alpha\beta + \beta\alpha)$ are possible during VQE optimization; the former is a multiconfigurational open shell singlet, the latter is the $S_z = 0$ component of the triplet. This also applies in the reverse, i.e. for de-excitations from the open-shell singlet, which explains why the VQE optimizer can sometimes yield a triplet state (see table \ref{spin_cross}) when initialized with an open singlet configuration (as in the case of CH$_2$ or NH$^{+}_2$ where the open shell singlet is the second excited state, separated from the triplet by a closed shell singlet), provided the required excitations are present in the cluster operator. This problem disappears when using $k$-UpCCGSD or UCCGD, as the double excitations contributing to spin crossover are removed from the wavefunction.

The results discussed for CH$_2$ and similar diradicals show the importance of single excitations in spin symmetry breaking during VQE optimization, and in \textit{k} convergence for \textit{k}-UpCC ans\"atze. To add further support to these findings, an approximation to the unitary form of Brueckner doubles coupled cluster (for an exact treatment see Mizukami \textit{et al} \cite{mizukami20}) is obtained for the CH$_2$ singlet by extracting the single excitations from an FCI calculation and forming a reference Brueckner determinant as described in section \ref{methods}. This was then used for the \textit{k}-UpCCGD and UCCGD operators. These give energy differences of below chemical accuracy (specifically, 0.1 mHa above and 0.5 mHa below the energy obtained for UCCGD and \textit{k}-UpCCGD, respectively) at the minimum energy bond length for both generalized doubles-only cluster operators, relative to the corresponding points shown in Figures \ref{ch2_curves_BL-only} and \ref{ch2_curves_singlets} for the S1 register. The $\tilde{a}^1A_1$ spin symmetry is also maintained during VQE optimization, as in the case for \textit{k}-UpCCGD and UCCGD. Thus, the role of single excitations in spin cross over during VQE optimization is confirmed, as is their role in \textit{k}-convergence in the \textit{k}-UpCC ans\"atze. 

In general, when VQE optimization is performed in the presence of multiple local minima near the optimal solution, careful examination of the initial state and its evolution during optimization is necessary, and possibly optimization constraints need to be imposed, for the algorithm to reach the correct energy surface. A representative case is the set of CH$_2$ calculations with the \textit{k}-UpCCGSD ansatz performed in this work. For the triplet state, small differences of approximately 1.5 mHa in the optimized energy were observed for \textit{k} = 1 as a function of initial cluster amplitude randomization. These energy differences are proportional to differences in the set of optimized cluster amplitudes, with the 1.5 mHa difference being accounted for by variations in amplitudes involving occupied-virtual and virtual-virtual double excitations. For the spread of 0.02 Ha in the energy of the ground state observed for calculations with \textit{k} $>$ 1, again virtual-virtual transitions play a role in the energy difference, however large variations in occupied-occupied single and double excitations are also observed, with the lowest energy solutions including occupied-occupied singles transitions which have very small amplitudes in the cases of higher total energy. The \% weights of each type of excitation for the triplet ground state within \textit{k}-UpCCGSD are shown in table \ref{table_amps}, where the lowest energy solutions from Figure \ref{ch2_Evsk_T} are used for each value of \textit{k}. The excitation weights are also shown for UCCGSD. The UCCGSD approach exhibits a large re-balancing of spectral weights relative to \textit{k}-UpCCGSD. In particular, the large weight of the generalized single excitations in 3-UpCCGSD are redistributed to the generalized doubles of UCCGSD. Note that the latter reproduces the FCI limit of the ground state energy to within 1 mHa, thanks to the inclusion of double excitations of spin orbitals in different spatial orbitals. 

This comparison explains the observed variation in accuracy of the \textit{k}-UpCCGSD approach applied to the ground state of CH$_{2}$; the greater degree of variational flexibility associated with larger values of \textit{k} allows for a partial compensation for the lack of generalized doubles in \textit{k}-UpCCGSD, resulting in a large transfer of spectral weight to the fully generalized singles (occupied-occupied transitions in this case) in order to achieve solutions closer to the FCI ground state. These results show how insufficient variational freedom in the excitation amplitudes (i.e. 1-UpCCGSD for $\tilde{X}^3B_1$), or neglecting generalized double excitations (1-UpCCGSD and 2-UpCCGSD for $\tilde{a}^1A_1$) in the cluster operator can result in inaccurate ground and excited state energies, respectively, for CH$_{2}$ and for complex molecules in general. 

\begin{table}
\caption{Optimized amplitudes for generalized excitations obtained with \textit{k}-UpCCGSD (\textit{k} = 1, 2, 3) and UCCGSD, for the triplet ($\tilde{X}^3B_1$) ground state of CH$_{2}$, categorized by single and double excitations. The weight of each excitation is given as a percentage of the total sum of coefficients in the cluster operator. Indices $A, B, C, D$ refer to occupied spatial orbitals, while $P, Q, R, S$ denote virtual spatial orbitals. For each value of \textit{k}, the lowest energy solution is used.}
\centering
\begin{tabular}{cccc} 
\hline
\textit{k} = 1 & & excitation & \% weight \\
\hline
\midrule
& \multirow{4}{10em}{\textit{singles}} & $t_{A}^{P}$ & 7.34 \\[5pt]
& & $t_{A}^{B}$ & 1.52 \\[5pt]
& & $t_{P}^{Q}$ & 1.19 \\
\midrule
& \multirow{4}{10em}{\textit{doubles}} & $t_{A,A}^{P,P}$ & 48.62 \\[5pt]
& & $t_{A,A}^{B,B}$ & 33.85 \\[5pt]
& & $t_{P,P}^{Q,Q}$ & 7.49 \\
\midrule
\textit{k} = 2 & & & \\
\hline
\midrule
& \multirow{4}{10em}{\textit{singles}} & $t_{A}^{P}$ & 10.73 \\[5pt]
& & $t_{A}^{B}$ & 29.68 \\[5pt]
& & $t_{P}^{Q}$ & 23.75 \\
\midrule
& \multirow{4}{10em}{\textit{doubles}} & $t_{A,A}^{P,P}$ & 10.90 \\[5pt]
& & $t_{A,A}^{B,B}$ & 12.86 \\[5pt]
& & $t_{P,P}^{Q,Q}$ & 12.08 \\
\midrule
\textit{k} = 3 & & & \\
\hline
\midrule
& \multirow{4}{10em}{\textit{singles}} & $t_{A}^{P}$ & 17.61 \\[5pt]
& & $t_{A}^{B}$ & 50.41 \\[5pt]
& & $t_{P}^{Q}$ & 14.20 \\
\midrule
& \multirow{4}{10em}{\textit{doubles}} & $t_{A,A}^{P,P}$ & 10.78 \\[5pt]
& & $t_{A,A}^{B,B}$ & 5.80 \\[5pt]
& & $t_{P,P}^{Q,Q}$ & 1.20 \\
\midrule
UCCGSD \\
\hline
\midrule
& \multirow{4}{10em}{\textit{singles}} & $t_{A}^{P}$ & 12.46 \\[5pt]
& & $t_{A}^{B}$ & 1.01 \\[5pt]
& & $t_{P}^{Q}$ & 0.52 \\
\midrule
& \multirow{8}{10em}{\textit{doubles}} & $t_{A,B}^{P,Q}$ & 37.35 \\[5pt]
& & $t_{A,B}^{C,D}$ & 1.01 \\[5pt]
& & $t_{P,Q}^{R,S}$ & 0.78 \\[5pt]
& & $t_{A,P}^{B,Q}$ & 3.79 \\[5pt]
& & $t_{A,B}^{C,P}$ & 26.85 \\[5pt]
& & $t_{A,P}^{Q,R}$ & 16.24 \\
\midrule
\end{tabular}
\label{table_amps}
\end{table}

\section{Conclusions}

In conclusion, VQE calculations using various ans\"{a}tze based on the unitary coupled cluster operator have been performed for the molecules H$_{2}$, H$_{3}$, H$_{4}$, NH, OH$^{+}$, CH$_{2}$, and NH$^{+}_{2}$, with different choices for the spin symmetry and number of electrons encoded in the qubit register. In the case of H$_{2}$ and H$_{3}$, the UCCSD energy curves obtained for each symmetry accurately match the expected results from calculations on classical machines at a similar level of theory.

The diradical molecules correspond to particularly interesting cases to assess the performance of different VQE ans\"{a}tze due to the presence of non-trivial contributions of electronic correlation to the ground and excited states. Using CH$_{2}$ to exemplify the main results observed for these systems, we find that while UCCSD, UCCGSD, and spin-paired 2-UpCCGSD all reproduce well the energy of the triplet ground state, the desired spin symmetry of the excited state can be maintained during VQE optimization by ans\"{a}tze which restrict the single excitation amplitudes to produce only singlet configurations. This reflects the argument put forward by ref. \cite{evangelista19} for commuting spin and cluster operators; this is confirmed in this work for realistic molecular states (beyond simplified toy models).
Efforts to use VQE to obtain the $\tilde{b}^1B_1$ state of CH$_{2}$ by setting the initial qubit register to an appropriate multireference open shell state fail for $k$-UpCCGS$^\prime$D and UCCGSD; however, \textit{k}-UpCCGSD with unrestricted singles manages to obtain a wavefunction of the correct symmetry for lower values of $k$, although with a significant overestimation of the energy associated to this state, and at higher values of $k$ the wavefunction is driven away from the open shell singlet in favor of the lower closed shell singlet. Removing single excitations and leaving only fully generalized doubles as in UCCGD can successfully access the open shell multireference singlet once a spin penalty is applied to the optimizer, indicating that single excitations can also be responsible for unwanted transitions between open and closed shell singlets during VQE optimization. 

Certain excited states can be accessed by VQE with the UCCSD ansatz without imposing penalty constraints in the objective function, as long as the parameters (qubit rotations applied to the initial qubit register) for mean-field state preparation are fixed during variational optimization of the cluster amplitudes. While fixing the mean-field state preparation can itself can be viewed as a constraint, this is a constraint that is imposed by knowledge of the electronic structure of the problem of interest. Hardware efficient ans\"{a}tze which provide increased variational freedom while improving computational efficiency need to be used with caution, as these ans\"{a}tze may introduce the possibility of different (undesired) Hilbert spaces being traversed during optimization. Similar care has to be taken when using generalized coupled cluster operators, especially on higher excited states when lower states of similar symmetry are present. Again, the dependency of the VQE algorithm on the initial configuration is emphasized. 

The results presented in this article shows that VQE can be used to calculate excited states. While well established for classical computing approaches to variational calculation of excited states, the ability of the VQE algorithm (combined with carefully adapted UCC ans\"atze) to capture strongly correlated multireference excites states of a range of molecules, without unwanted breaking of spin or particle number symmetry, is now demonstrated in this work. In particular, once the excitations responsible for unwanted symmetry transitions are identified, these can be removed, which also yields the added benefit of a smaller number of variational parameters. This work also shows the importance of testing VQE ans\"{a}tze on molecules or other systems with sufficient complexity, preferably beyond simple toy models. In particular, for studying many-body electronic systems using different approximations developed to balance efficiency with accuracy, clearly only systems with sufficient electronic correlation can test the full extent of the (in)adequacies of these approximations. 

\section{Acknowledgements}

We would like to acknowledge useful discussions with Dr. Yu-ya Ohnishi from JSR Corporation (Japan).

\printbibliography
\end{document}